\documentclass[aps,prd,
twocolumn,
amsmath,amssymb,
groupedaddress
]{revtex4-2}

\usepackage{hyperref}
\usepackage{multirow}
\hypersetup{
    colorlinks=true,
    linkcolor=blue,
    filecolor=blue,      
    urlcolor=blue,
    citecolor=blue}
\begin{document}

%\preprint{}

\title{Jordan and Einstein Frames from the perspective of $\omega=-3/2$ Hamiltonian Brans-Dicke theory}



\author{Matteo Galaverni}
\email[]{matteo.galaverni@gmail.com}
\affiliation{Specola Vaticana (Vatican Observatory), V-00120 Vatican City, Vatican City State}
\affiliation{INAF/OAS Bologna, via Gobetti 101, I-40129 Bologna, Italy\\
ORCiD: 0000-0002-5247-9733}

\author{Gabriele Gionti, S.J.}
\email[Corresponding author: ]{ggionti@specola.va}
%\thanks{}
\affiliation{Specola Vaticana (Vatican Observatory), V-00120 Vatican City, Vatican City State}
\affiliation{Vatican Observatory Research Group, Steward Observatory, The University Of Arizona,
933 North Cherry Avenue, Tucson, Arizona 85721, USA}
\affiliation{INFN, Laboratori Nazionali di Frascati, Via E. Fermi 40, I-00044 Frascati, Italy.\\
ORCiD: 0000-0002-0424-0648}


\date{\today}

\begin{abstract}
We carefully perform a Hamiltonian Dirac's constraint analysis of $\omega=-\frac{3}{2}$ Brans-Dicke theory with Gibbons-Hawking-York (GHY) boundary term. The Poisson brackets are computed via functional derivatives. After a brief summary of the results for $\omega\neq-\frac{3}{2}$ case  \cite{Gionti2021}, we derive all Hamiltonian Dirac's constraints and constraint algebra both in the Jordan and Einstein frames. Confronting and contrasting Dirac's constraint algebra in both frames, it is shown that they are not equivalent. This highlights  the transformations from the Jordan to the Einstein frames are not Hamiltonian canonical transformations.    
\end{abstract}

\maketitle

\section{\label{Jordan-Einstein} Introduction}

We consider a scalar-tensor theory action \cite{Dyer}  with the Gibbons-Hawking-York (GHY) boundary term \cite{gibbons&hawking} \cite{york1} \cite{york2}
\begin{eqnarray}
S&=&\int_{M}d^{n}x{\sqrt{-g}}\left(f(\phi)R-\frac{1}{2}\lambda(\phi)g^{\mu\nu}\partial_{\mu}\phi\partial_{\nu}\phi -U(\phi)\right) \nonumber \\
&&+2\int_{\partial M}d^{n-1}{\sqrt{h}}f(\phi)K \, ,
\label{scalartensor}
\end{eqnarray}
where $f(\phi)$ is a generic function of $\phi$ as well as $\lambda(\phi)$, $K$ is the trace of the extrinsic curvature. Varying the previous action respect to the metric $g^{\mu\nu}$ with the condition that on the boundary the variation of it be zero $\delta g^{\mu\nu}=0$, we obtain the equation of General Relativity for the case of a scalar-tensor theory of Gravity
\begin{equation}
f(\phi)\left(R_{\mu\nu}-\frac{1}{2}g_{\mu\nu}R\right)+g_{\mu\nu}\Box f(\phi)-\nabla_{\mu}\nabla_{\nu}f(\phi)=T^{\phi}_{\mu\nu}\,, 
\label{Einsteinequiv}
\end{equation}
where 
\begin{equation}
T^{\phi}_{\mu\nu}=\frac{\lambda(\phi)}{2}\left(\partial_{\mu} \phi \partial _{\nu} \phi -\frac{1}{2}g_{\mu\nu}g^{\alpha\beta}\partial_{\alpha} \phi \partial _{\beta} \phi \right)-\frac{1}{2}g_{\mu\nu}U(\phi)\,.
\label{tensorimpu}
\end{equation}

Varying respect to $\phi$ and imposing that its variations on the boundary is zero as well, $\delta \phi =0$, the equation of evolution for $\phi$ is 
\begin{equation}
f'(\phi)R+\frac{1}{2}\lambda'(\phi)(\partial \phi)^{2}+\lambda(\phi) \Box \phi -U'(\phi)=0\,.
\label{eqaphi}
\end{equation}

Nowadays, one says to pass from the {\it Jordan}, where the action is a scalar-tensor theory \eqref{scalartensor}, to the {\it Einstein} frame \cite{Dicke} \cite{Faraoni2006} through a Weyl (conformal) transformation of the metric tensor, keeping the scalar field $\phi(x)$ unchanged,
\begin{eqnarray}
{\widetilde g}_{\mu\nu}&=&\Big(16\pi G f(\phi)\Big)^{\frac{2}{n-2}}g_{\mu\nu}\,,\nonumber\\
{\widetilde \phi}(x)&=&\phi(x)\,,
\label{Weyltrans1}
\end{eqnarray}
${\widetilde g}_{\mu\nu}$ and ${\widetilde \phi}(x)$ being the transformed metric tensor and scalar field.
In the {\it Einstein frame} the action \eqref{scalartensor} becomes 
\begin{eqnarray}
S&=&\int_{M}d^{n}x{\sqrt{-{\widetilde g}}}\left(\frac{1}{16\pi G}{\widetilde R}-A(\phi){\widetilde g}^{\mu\nu}\partial_{\mu}\phi\partial_{\nu}\phi -V(\phi)\right) \nonumber \\
&&+\frac{1}{8\pi G}\int_{\partial M}d^{n-1}{\sqrt{\widetilde h}}{\widetilde K}
\label{scalartensorEF},
\end{eqnarray}
where 
\begin{eqnarray}
A(\phi)&=&\frac{1}{16\pi G}\left(\frac{\lambda(\phi)}{2f(\phi)}+\frac{n-1}{n-2}\frac{(f'(\phi))^2}{f^2(\phi)}\right), \nonumber \\ 
V(\phi)&=&\frac{U(\phi)}{[16\pi G f(\phi)]^{\frac{n}{n-2}}}.
\label{AandV}
\end{eqnarray}
varying this equation respect to ${\widetilde g}^{\mu\nu}$ we get Einstein equations and varying respect to $\phi$ we get the equation for $\phi(x)$. 

In literature, \cite{Dicke} \cite{Faraoni2006} when $(g_{\mu\nu}(x),\phi(x))$ is a solution of the equations of motion in the Jordan frame, the corresponding couple $(\widetilde{g}_{\mu\nu}(x,\phi), \phi(x))$ is assumed to be solution of the equations of motion in the Einstein frame. In this way, the two frames are considered physically equivalent. In fact one is doing nothing else but imposing the same solutions, linked by a Weyl (conformal) transformation, in the two frames.  

The physics behind this transformation dates back to an idea of Dicke. He, in a seminal paper \cite{Dicke}, observed that physics is invariant under re-definition of unit of measurements. 
If we re-scale the length unit by a factor $\lambda$ such that the value of the square of the line element, in the new unit, is $d{\widetilde s}^{2}={\lambda}^{2}d{s}^{2}$ (recall that the definition of the line element is $ds^{2}\equiv g_{\mu\nu}dx^{\mu}dx^{\nu}$), the relation between the metric coefficients under unit length re-scaling is  ${\widetilde g}_{\mu\nu}={\lambda}^{2} g_{\mu\nu}$. Therefore invariance of the physical observables under rescaling of units of measurements implies invariance under Weyl rescaling of the metric tensor \cite{Dicke}.

Physical equivalence of the observable quantities in the Jordan and in the Einstein frames have been very much debated \cite{Capozziello2010} \cite{Deruelle2010h} \cite{Cho1992} \cite{Nandi1} \cite{Nandi2} \cite{Nandi3} \cite{fluid1} \cite{fluid2} \cite{fluid3} \cite{Carloni:2010rfq}. On average, the community seems to be in favour of the physical equivalence although the interpretation of the experiments might be different \cite{Francfort2019}. We personally think that the equivalence works mathematically as long as one is sure the solutions can be mapped from one frame to the other, although some mathematical concerns have been raised as well \cite{Kamenshchik:2016gcy}. 

We continue to study the Hamiltonian canonical equivalence between the two frames started in the article \cite{Gionti2021}. Now on, we analyze the question of the equivalence using the Hamiltonian Dirac's constraint analysis for the particular value of the Brans-Dicke parameter $\omega=-\frac{3}{2}$.

We will summarize in Section~\ref{bdicke} the results of the Hamiltonian analysis of the Brans-Dicke theory for $\omega \neq -\frac{3}{2}$ and the (Hamiltonian) canonical in-equivalence between the two frames for this case \eqref{comparison}. Section~\ref{Section3} deals with the Hamiltonian analysis of the case $\omega=-\frac{3}{2}$: after having shown the extra Weyl (conformal) symmetry of the action for this case \eqref{Conformal21}, we perform the ADM decomposition \eqref{ADMpart} and study the constraint algebra among the Dirac's Hamiltonian constraints \eqref{Hamiltonian Constraint}; $\omega=-\frac{3}{2}$ Hamiltonian Brans-Dicke theory is examined in the Einstein frame \eqref{JF-EF}, and final remark, regarding the different Dirac's Hamiltonian constraint algebra, is addressed in \ref{canonical2}.
We conclude in Section~\ref{Conclusions}.

\section{\label{bdicke}Brans-Dicke theory for $\omega \neq -\frac{3}{2}$  }
In recent years, much research has been done to study the classical Hamiltonian equivalence between Jordan and Einstein frames \cite{Barreto2017} \cite{Deruelle2009} \cite{Ezawa1998} \cite{Ezawa2009} as well as their quantum equivalence \cite{Falls2018} \cite{Kamenshchik2014} \cite{Ohta2017} \cite{Filippo2013}
\cite{Frion2018}.
We summarize here the results of Dirac's constraint Hamiltonian analysis \cite{dirac1966} \cite{Esposito1992} \cite{Cawley}
(see also \cite{Olmo} \cite{Gielen} \cite{floreaniniJackiw} \cite{costagirotti} \cite{faddeevJackiw} \cite{Kiefer2017} \cite{Barvinsky2019} for complementary cases) of Brans-Dicke theory in the two frames \cite{Gionti2021}. 

Brans-Dicke theory \cite{Brans1961} is a special case of \eqref{scalartensor} when $f(\phi)=\phi$ and $\lambda(\phi)=\frac{2\omega}{\phi}$ \cite{Dyer}:
\begin{eqnarray}
S&=&\int_{M}d^{4}x\sqrt{-g}\left(\phi\;{}^{4}R-\frac{\omega}{\phi}g^{\mu\nu}\partial_{\mu}\phi \partial_{\nu} \phi -U(\phi)\right)\nonumber \\
&&+ 2\int_{\partial M} d^3x \sqrt{h}\phi K\;.
\label{BDaction}
\end{eqnarray}

The equations of motion for the metric tensor $g_{\mu\nu}$ are a particular case of \eqref{Einsteinequiv}
\begin{eqnarray}
R_{\mu \nu }&-&\frac{1}{2}g_{\mu \nu }R=\frac{\omega}{\phi^2}\left[\partial_\mu\phi\partial_\nu\phi-\frac{1}{2}g_{\mu\nu}g^{\alpha\beta}\partial_\alpha
 \phi \partial_\beta \phi)\right]\nonumber\\
 &+&\frac{1}{\phi}\left[\nabla_\mu\nabla_\nu\phi-g_{\mu\nu}\Box \phi -\frac{1}{2}g_{\mu\nu}U(\phi)\right]
 \label{equationforg},
\end{eqnarray}
while the equation of motion for $\phi$, a particular case of \eqref{eqaphi}, is 
\begin{equation}
(3+2\omega)\Box \phi =\phi \frac{dU}{d\phi}-2U(\phi)\,.
\label{eqstophi}
\end{equation}

\subsection{\label{ADM0part}ADM decomposition and definition of the ADM Hamiltonian density function}

The Arnowitt-Deser-Misner (ADM)-decomposition \cite{ADM} is based on the assumption that the topology of the Space-Time $(M,g)$ is $M={\mathbb{R}}\times \Sigma$ \cite{Esposito1992}; where $\mathbb{R}$ is a one dimensional space, the time direction, 
and $\Sigma$ is a three dimensional space-like surface embedded in $M$. $g$ is the ADM-metric tensor defined as 
\begin{eqnarray}
g&=&-(N^{2}-N_{i}N^{i})dt \otimes dt +N_{i}(dx^{i} \otimes dt
+dt \otimes dx^{i})\nonumber\\
&&+h_{ij}dx^{i} \otimes dx^{j}
\label{ADMmetricJF}\,,
\end{eqnarray}
where $N=N(t,x)$ is the lapse function, $N^{i}=N^{i}(t,x)$ are the shift functions \cite{DeWitt1967} \cite{Esposito1992} \cite{Gionti2021}.

The ADM Lagrangian density $\mathcal{L}_{ADM}$ is
\begin{eqnarray}
\label{eq:Lagrangian2}
\mathcal{L}_{ADM}&=& {\sqrt{h}} \Bigg[N \phi\left( {}^{(3)}R+K_{ij}K^{ij}-K^2\right)\nonumber \\
&&-\frac{\omega}{N\phi}\left(N^2 h^{ij}D_i\phi D_j\phi- (\dot{\phi}-N^i D_i\phi)^2\right)  \\
&&+2K (\dot{\phi}-N^iD_i\phi )-NU(\phi)+2h^{ij}D_iN D_j\phi\Bigg]\;, \nonumber
\label{scompi}
\end{eqnarray}
 $K_{ij}$ is the extrinsic curvature defined as follows \cite{Esposito1992}
\begin{equation}
{K}_{ij}=\frac{1}{2 N}
\left(-\frac{\partial h_{ij}}{\partial t} +D_{i}{N}_{j} +D_{j}{N}_{i}\right),
\label{lextrins}
\end{equation}
The canonical momenta $(\pi_N, \pi_{i}, \pi^{ij}, \pi_{\phi})$ associated  to $(N, N^{i}, h_{ij}, \phi)$ are  
\begin{eqnarray}
\pi_N &=&\frac{\partial {\mathcal L}_{ADM}}{\partial \dot{N}}\approx 0 \ , \pi_i=\frac{\partial {\mathcal L}_{ADM}}{\partial \dot{N}^i}\approx 0 \ ,    \nonumber \\
\pi^{ij}&=&\frac{\partial {\mathcal L}_{ADM} }{\partial \dot{h}_{ij}}=-{\sqrt{h}}\left[ \phi \Big(K^{ij}-Kh^{ij}\Big)+\frac{h^{ij}}{N}\Big(\dot{\phi}-N^iD_i\phi\Big)\right] \label{pippo1}   \ , \ \nonumber \\ 
\pi_\phi&=&\frac{\partial {\mathcal L}_{ADM}}{\partial \dot{\phi}}={\sqrt{h}}\left( 2K+\frac{2\omega}{N\phi}(\dot{\phi}-N^iD_i\phi)\right)\;,
\end{eqnarray}
which show the momenta $\pi_N$ and $\pi_i$ are primary constraints ($\approx$ meaning the quantity is zero on the constraint surface) \cite{dirac1966} \cite{Esposito1992} \cite{Menotti2017}. 

The hamiltonian density $\mathcal{H}_{ADM}$ is  
\begin{equation}
{\mathcal H}_{ADM}={\pi}^{ij}{\dot {h}}_{ij}+{\pi}_{\phi}{\dot \phi}-\mathcal{L}_{ADM}\;.
\label{hamiltodefin}
\end{equation}
This definition holds on the constraint surface defined by the Dirac's primary constraints $\pi_N \approx 0 $ and $\pi^{i}\approx 0 $ found above \eqref{pippo1} \cite{dirac1966} \cite{Esposito1992}.
The Hamiltonian density ${\mathcal H}_{ADM}$ is
\begin{eqnarray}
{\mathcal{H}}_{ADM}&=&{\sqrt{h}}\Bigg\{ N\left[-\phi\; {}^{3}R+\frac{1}{\phi h}\left( \pi^{ij}\pi_{ij}-\frac{{\pi_h}^2}{2}\right)\right] \nonumber \\
&&+ \frac{N\omega}{\phi}D_i\phi D^i\phi 
+N2D^iD_i\phi + NV(\phi) \nonumber  \\
 &&+\frac{1}{2 h\phi}\left(\frac{N}{3+2 \omega}\right)
 (\pi_h - \phi\pi_{\phi})^2\Bigg\}\nonumber\\
 && -2N^iD_j\pi^{j}_{i}+N^iD_i\phi \pi_{\phi}\;, \label{hamiltoeff1}
\end{eqnarray}
where $\pi_h\equiv \pi^{ij}h_{ij}$,
and it can be written in the following form 
\begin{equation}
{\mathcal{H}}_{ADM}=N{\mathcal H}+N^{i}{\mathcal H}_{i}\,,
\label{scompositio1}
\end{equation}
where the $\mathcal H$ is the Hamiltonian density constraint
\begin{eqnarray}
\mathcal{H}&=&\left[-\phi\;  {}^{3}R+\frac{1}{\phi h}\left( \pi^{ij}\pi_{ij}-\frac{{\pi_h}^2}{2}\right)\right] \nonumber\\
&+& \frac{\omega}{\phi}D_i\phi D^i\phi 
+2D^iD_i\phi                                     
 +V(\phi)\nonumber\\ 
 &&+\frac{1}{2 h\phi}\left(\frac{1}{3+2 \omega}\right)
 (\pi_h - \phi\pi_{\phi})^2\,,
\end{eqnarray}
and ${\mathcal {H}}_i$ is the momentum constraint 
\begin{equation}
{\mathcal {H}}_i= -2D_j\pi^{j}_{i}+D_i\phi \pi_{\phi}\;. 
\label{momentumcons1}
\end{equation}

Therefore the total Hamiltonian $H_{T}$ \cite{Esposito1992} is 
\begin{equation}
H_{T}=\int d^{3}x \left(\lambda \pi_N + \lambda^{i}\pi_{i}+N{\mathcal{H}}+N^i{\mathcal{H}}_{i} \right)\,, 
\label{hamiltonianatot1}
\end{equation}
$\lambda=\lambda(t,x)$ and $\lambda^{i}(t,x)$ being Lagrange multipliers.

If we indicate the canonical variables $(N,N^i, h_{ij}, \pi_N, \pi_{i}, \pi^{ij})$ generically with $(Q^i,\Pi_i)$ the Poisson Brackets between two arbitrary function $A$ and $B$ of the canonical variables are 
\cite{Menotti2017} 
\begin{equation}
\left\{A,B\right\}=\int d^3y \left(\frac{\delta A}{\delta Q^{i}(y)}\frac{\delta B}{\delta \Pi_{i}(y)}-\frac{\delta A}{\delta \Pi_{i}(y)}\frac{\delta B}{\delta Q^{i}(y)}\right).
\label{PoissonBra}
\end{equation}
In reference \cite{Gionti2021} it has been shown that the algebra of the secondary constraints is like Einstein's geometro-dynamics
\begin{eqnarray}
 \left\{{\mathcal H}_i(x), {\mathcal H}_j(x')\right\} &=&{\mathcal H}_i(x') \partial_j \delta(x,x')- {\mathcal H}_j(x) {\partial_i}' \delta(x,x')\,,\nonumber \\  
 \left\{{\mathcal H}(x), {\mathcal H}_i(x')\right\}&=&-{\mathcal H}(x'){{\partial}'_i}\delta(x,x')\,, \label{differential} \\
\{{\cal{H}}(x),{\cal{H}}(x')\}&=&{\cal {H}}^{i}(x)\partial_{i}\delta(x,x')-{\cal H}^{i}(x'){\partial}'_{i}\delta(x,x')\,. \nonumber
\end{eqnarray}
As extensively argued in \cite{Kuchar2} and \cite{Kuchar1}, once matter source is introduced with its own canonical variables, many different inequivalent theories of gravity coupled with matter can generate the same constraint algebra \eqref{differential}.

\subsection{\label{comparison}Transformations from the Jordan to the Einstein Frame }

The Weyl (conformal) transformation \eqref{Weyltrans1} entails an ADM metric in the Einstein frame, cfr.  \cite{Deruelle2009} \cite{Garay1992},
\begin{eqnarray}
\widetilde {g}&=&-({\widetilde N}^{2}-{\widetilde N}_{i}\widetilde{N}^{i})dt \otimes dt 
+\widetilde{N}_{i}(dx^{i} \otimes dt
+dt \otimes dx^{i})\nonumber\\
&&+\widetilde{h}_{ij}dx^{i} \otimes dx^{j}\;, 
\label{EFmetricADM1}
\end{eqnarray}
where
\begin{eqnarray}
&&\widetilde{N}=\left(16\pi G f(\phi) \right)^{\frac{1}{n-2}}N\,, \widetilde{N}_i=\left(16\pi G f(\phi) \right)^{\frac{2}{n-2}}N_i\,,\nonumber\\
&&\widetilde{h}_{ij}=\left(16\pi G f(\phi) \right)^{\frac{2}{n-2}}h_{ij}.
\label{tilderelation0}
\end{eqnarray}

The canonical momenta in the Einstein Frame, associated to the variables $\eqref{tilderelation0}$,
in the $\omega\neq-\frac{3}{2}$ Brans-Dicke case, are (cfr. \cite{Gionti2021})   
\begin{eqnarray}
\label{momentumEF}
&&{\widetilde \pi}^{ij}= \frac{\partial {\mathcal {\widetilde L}}_{ADM} }{\partial \dot{\widetilde{h}}_{ij}}=
-\frac{\sqrt{\widetilde {h}}}{{16 \pi G}}\left( {\widetilde K}^{ij}-{\widetilde K}{\widetilde h}^{ij}\right)=\frac{{\pi}^{ij}}{16\pi G\phi}\,,\\
&&{\widetilde \pi}_\phi=\frac{\partial {\mathcal {\widetilde L}}_{ADM}}{\partial \dot{\phi}}=\frac{\sqrt{\widetilde {h}}(\omega +\frac{3}{2})}{8\pi G {\widetilde N}{\phi}^2}\left(\dot{\phi}-{\widetilde N}^i\partial_i\phi \right)
%\nonumber \\&&
=\frac{1}{\phi}(\phi \pi_{\phi}-\pi_{h})\,.\nonumber
\end{eqnarray}

The transformations \eqref{tilderelation0} \eqref{momentumEF} are assumed a canonical set, in Hamiltonian sense, of variables \cite{Deruelle2009}. But this is not completely true (cfr. \cite{Kiefer2017}) since,
see also \cite{Gionti2021} for more details
\begin{equation}
\{{\widetilde N},{\widetilde \pi}_{\phi} \}=\frac{8\pi GN}{{\sqrt{16 \pi G\phi}}} \neq 0, \textrm{ and }\; 
\{{\widetilde N}_i,{\widetilde \pi}_{\phi} \}=16 \pi G N_i \neq 0\,,
\label{noncanonicalcond}
\end{equation}
where, obviously, the Poisson brackets are calculated in the Jordan Frame. Therefore, since the transformations from the Jordan to the Einstein frames are not canonical, one is not allowed to pass from the Jordan to the Einstein frame to perform the constraint analysis of the Brans-Dicke theory as it is usually done (cfr. \cite{Garay1992}). 

The Hamiltonian canonical transformations \cite{Gionti2021} hold the lapse and the shifts ${\widetilde {N^{*}}}=N$ and ${\widetilde {N^{*}}}^{i}=N^{i}$ while $h_{ij}$ and $\phi$ and their respective momenta $\pi^{ij}$ and $\pi_{\phi}$ transform according to the equations \eqref{tilderelation0} and \eqref{momentumEF}. These transformations generate an Anti-Newtonian Gravity as explained in \cite{Niedermaier2019} \cite{Niedermaier2020} \cite{Zhang2011} \cite{Zhou2012}. They correspond to the following scaling relation on the ADM metric
\begin{eqnarray}
N\mapsto N\,; N^{i}\mapsto N^{i}\,; h_{ij}\mapsto \lambda^{2}h_{ij}\,,
\label{tredimesconf}
\end{eqnarray}
and for $\lambda \gg 1$ emulate a large value of the Newton constant $G$ and also enhance space-like distances over to time-like ones \cite{Niedermaier2019} \cite{Niedermaier2020}. In practice the Weyl (Conformal) transformation is implemented only on the three-dimensional metric $h_{ij}$ of the three-dimensional space-surfaces $\Sigma$.

In ref. \cite{Gionti2021}, we considered the finite dimensional case of a  mini-superspace model built from the action \eqref{BDaction} evaluated on a flat FLRW metric. The correspondent set of transformations from the Jordan to the Einstein frame, analogous to \eqref{tilderelation0} \eqref{momentumEF} for this particular case, still shows to be not canonical (in the Hamiltonian sense). The Anti-Newtonian Gravity transformations \eqref{tredimesconf} represent still the canonical one's.

\section{\label{Section3} Brans-Dicke theory for the case $\omega=-\frac{3}{2}$}

In this section we study in detail the particular case of Brans-Dicke action \eqref{BDaction} for $\omega=-\frac{3}{2}$
\begin{eqnarray}
S^{(-3/2)}&=&\int_{M}d^{4}x\sqrt{-g}\left(\phi R+\frac{3}{2}\frac{g^{\mu\nu}}{\phi}\partial_{\mu}\phi \partial_{\nu} \phi -U(\phi)\right)\nonumber \\
&&+ 2\int_{\partial M} d^3x \sqrt{h}\phi K\;.
\label{BDaction2}
\end{eqnarray}
We introduced the superscript $\cdots^{(-3/2)}$ in order to underline when a 
quantity is evaluated in the particular case $\omega=-\frac{3}{2}$, see also \cite{Zhang2011} \cite{Gielen}. $M$ is a manifold with a boundary $\partial M$ on which it is defined the three-metric $h$, pull-back of $g$ on the boundary $\partial M$ and the extrinsic curvature $K_{ij}$. The potential $U(\phi)$ is the same as in Eq. \eqref{BDaction} and is always, for consistency reasons from equation \eqref{eqstophi}, of the form $\alpha {\phi}^{2}$, where $\alpha$ is a generic constant.
The action \eqref{BDaction2} is made out of two terms: a bulk term $S_{M}^{(-3/2)}$
\begin{equation}
S_{M}^{(-3/2)}=\int_{M}d^{4}x\sqrt{-g}\left(\phi R+\frac{3}{2}\frac{g^{\mu\nu}}{\phi}\partial_{\mu}\phi \partial_{\nu} \phi -U(\phi)\right)\,,
\label{primo}
\end{equation}
and a boundary term $S_{\partial M}^{(-3/2)}$
\begin{equation}
S_{\partial M}^{(-3/2)}=
 2\int_{\partial M} d^3x \sqrt{h}\phi K\;.
\label{BBDaction}
\end{equation}

\subsection{\label{Conformal21} (Invariance under) Conformal Transformations}

If we perform the following Weyl (conformal) transformation
\begin{equation}
{\widetilde g}_{\mu\nu}={\Omega}^{2}g_{\mu\nu}\,,
\label{conformal}
\end{equation}
the trace $R$ of the Ricci tensor $R_{\mu\nu}$ transforms in the following way \cite{Dabrowski2008}
\begin{equation}
R={\Omega}^{2}\left({\widetilde R}+ \frac{6\,{\widetilde{\Box}\Omega}}{\Omega}
-12{\widetilde g}^{\mu\nu}\frac{{\Omega}_{,\mu}{\Omega}_{,\nu}}{{\Omega}^{2}}\right)\,.
\label{Rconfo}
\end{equation}
The Weyl (conformal) transformation on the field $\phi$ is 
\begin{equation}
{\widetilde {\phi}}=\frac{\phi}{{\Omega}^2}
\label{campoconfo}\,.
\end{equation}
If we apply the previous Weyl (conformal) transformations on the action $S^{(-3/2)}=S_{M}^{(-3/2)}+S_{\partial M}^{(-3/2)}$, we get on $S_M^{(-3/2)}$
\begin{eqnarray}
S_{M}^{(-3/2)}&=&\int_{M}d^{4}x\sqrt{-{\widetilde{g}}}\left\{\big[{\widetilde{\phi}} {\widetilde R}+ \frac{6\,{\widetilde{\Box}}\Omega}{\Omega}\widetilde{\phi}
-12{\widetilde g}^{\mu\nu}\frac{{\Omega}_{,\mu}{\Omega}_{,\nu}}{{\Omega}^2}{\widetilde{\phi}}\big]\right.\nonumber\\
&&\left. + \frac{3}{2}\frac{{\widetilde{g}}^{\mu\nu}}{{\widetilde{\phi}}}\partial_{\mu}{\widetilde{\phi}}\partial_{\nu}{\widetilde{\phi}}
+\frac{6}{\Omega}{\widetilde{g}}^{\mu\nu}\partial_{\mu}{\Omega}\partial_{\nu}{\widetilde{\phi}}\right.\nonumber\\
&&\left.
+\frac{6}{\Omega^2}{\widetilde{\phi}}\, {\widetilde{g}}^{\mu\nu}
\partial_{\mu}{\Omega\partial_{\nu}}{\Omega}
-\alpha{\widetilde{\phi}}^{2}\right\}\, ,
\label{rompe}
\end{eqnarray}
while on $S_{\partial M}^{(-3/2)}$ 
\begin{equation}
S_{\partial M}^{(-3/2)}=2\int_{\partial M}d^{3}x \sqrt{{\widetilde{h}}}{\widetilde \phi}{\widetilde K}-6\int_{\partial M}d^{3}x \sqrt{{\widetilde{h}}}\frac{\widetilde{\phi}}{\Omega}
{\widetilde{n}}^{\mu}{\nabla}_{\mu}\Omega\,\, ,
\label{conformalboundary}
\end{equation}

here $\widetilde{K}$ is the Weyl (conformal) transformed  extrinsic curvature, ${\widetilde{n}}^{\mu}$ the Weyl (conformal) transformed normal vector to the boundary $\partial M$, as can be found in the appendix of \cite{Dyer}.

The Weyl (conformal) transformation on the full action $S^{(-3/2)}$, simplifying the previous expression, is 
\begin{eqnarray}
S^{(-3/2)}&=&S_{M}^{(-3/2)}+S_{\partial M}^{(-3/2)}\nonumber\\
&=&\int_{M}d^{4}x\sqrt{-{\widetilde{g}}}\left({\widetilde{\phi}} {\widetilde R}+\frac{3}{2}\frac{{\widetilde{g}}^{\mu\nu}}{{\widetilde{\phi}}}\partial_{\mu}{\widetilde{\phi}}\partial_{\nu}{\widetilde{\phi}}-
U(\widetilde{\phi})\right)\nonumber\\
&&+2\int_{\partial M}d^{3}x \sqrt{{\widetilde{h}}}{\widetilde \phi}{\widetilde K}\,.
\label{finale}
\end{eqnarray}
This proves that Brans-Dicke  \eqref{BDaction2} is invariant under Weyl (conformal) transformations \eqref{conformal}- \eqref{campoconfo}
in the particular case $\omega=-\frac{3}{2}$.

\subsection{\label{ADMpart}ADM decomposition and definition of the ADM Hamiltonian density function}
The ADM Brans-Dicke Lagrangian density ${\mathcal L}_{ADM}$, introduced in Eq. \eqref{eq:Lagrangian2} for a generic $\omega$,
is here specified for $\omega=-\frac{3}{2}$
\begin{eqnarray}\label{eqLagrangian2}
\mathcal{L}_{ADM}^{(-3/2)}&=& {\sqrt{h}} \Bigg[N \phi\left( {}^{(3)}R+K_{ij}K^{ij}-K^2\right)\nonumber \\
&&+\frac{3}{2N\phi}\left(N^2 h^{ij}D_i\phi D_j\phi- (\dot{\phi}-N^i D_i\phi)^2\right)  \\
&&+2K (\dot{\phi}-N^iD_i\phi )-NU(\phi)-2h^{ij}N D_iD_j\phi\Bigg]\;. \nonumber
\end{eqnarray}
In this case, the canonical momenta $(\pi_N, \pi_{i}, \pi^{ij}, \pi_{\phi})$  associated  to $(N, N^{i}, h_{ij}, \phi)$ are obtained from \eqref{pippo1} for $\omega=-\frac{3}{2}$. 
As in the general case $\omega \neq -\frac{3}{2}$, the momenta $\pi_N$ and $\pi_{i}$ associated to the lapse $N$ and the shifts $N^{i}$ are primary constraint. 
An extra primary constraints, consequence of the Weyl (conformal) symmetry discussed above, is 

\begin{equation}
C_{\phi}\equiv \pi^{ij}h_{ij}-\phi\pi_{\phi}\approx 0\,\,,
\label{extraprimary}
\end{equation}
see Eq.~\eqref{momentumEF}, we name it conformal constraint. 

The ADM-Hamiltonian ${\mathcal{H}_{ADM}}$ is, as usual, defined in the following way, see \eqref{hamiltodefin}
\begin{equation}
{\mathcal H}_{ADM}={\pi}^{ij}{\dot {h}}_{ij}+{\pi}_{\phi}{\dot \phi}-\mathcal{L}_{ADM}\;.
\label{hamiltodefin2}
\end{equation}
The explicit form is
\begin{eqnarray}
{\mathcal{H}}_{ADM}^{(-3/2)}&=&{\sqrt{h}}\Bigg\{ N\left[-\phi\;  {}^{3}R+\frac{1}{\phi h}\left( \pi^{ij}\pi_{ij}-\frac{{\pi_h}^2}{2}\right)\right] \nonumber \\
&&- \frac{3N}{2\phi}D_i\phi D^i\phi 
+N2D^iD_i\phi                                     
 +NU(\phi) \label{hamiltoeff2} \Bigg\}\nonumber\\
&& -2N^iD_j\pi^{j}_{i}+N^iD_i\phi \pi_{\phi}\;,  
 \label{HamiltoniandensityDBJF}
\end{eqnarray}
and can be re-written in the following form 
\begin{equation}
{\mathcal{H}}_{ADM}^{(-3/2)}=N{\mathcal H}^{(-3/2)}+N^{i}{\mathcal H}_{i}^{(-3/2)},
\label{scompositio2}
\end{equation}
where the $\mathcal H^{(-3/2)}$ is the Hamiltonian constraint, 
and is just: 
\begin{eqnarray}
{\mathcal H}^{(-3/2)}&=&{\sqrt{h}}\Bigg\{ \left[-\phi\;  {}^{3}R+\frac{1}{\phi h}\left( \pi^{ij}\pi_{ij}-\frac{{\pi_h}^2}{2}\right)\right] \nonumber\\
&&- \frac{3}{2\phi}D_i\phi D^i\phi+2D^iD_i\phi+U(\phi) \Bigg\}\,,
 \label{hamiltoconst}
\end{eqnarray}
and ${\mathcal {H}}_i^{(-3/2)}$ is the momentum constraints 
\begin{equation}
{\mathcal {H}}_i^{(-3/2)}= -2D_j\pi^{j}_{i}+D_i\phi \pi_{\phi}\;. 
\label{momentumcons2}
\end{equation}

The total Hamiltonian $H_{T}^{(-3/2)}$ \cite{Esposito1992} is
\begin{eqnarray}
H_{T}^{(-3/2)}&=&\int d^{3}x \left(\lambda \pi_N + \lambda^{i}\pi_{i} + \lambda_{\phi}C_{\phi}\right.\nonumber\\
&&\left.+N{\mathcal{H}^{(-3/2)}}+N^i{\mathcal{H}}_{i}^{(-3/2)} \right)\;\,, 
\label{hamiltonianatot2}
\end{eqnarray}
where $\lambda=\lambda(t,x)$, $\lambda^{i}(t,x)$, and $\lambda_{\phi}(t,x)$ are Lagrange multipliers. 

The preservation of the primary constraints $\pi_N \approx 0$ and $\pi_i \approx 0$ along the dynamic generated by the total Hamiltonian $H_T^{(-3/2)}$ \eqref{hamiltonianatot2} gives
\begin{equation}
{\dot \pi}_N=\{\pi_N, H_T^{(-3/2)}\}=-{\mathcal H^{(-3/2)}} \approx 0 \, ,
\label{hamiltonianconstro}
\end{equation}
and 
\begin{equation}
{\dot {\pi}_{i}}=\{\pi_i, H_T^{(-3/2)}\}=-{\mathcal H}_i^{(-3/2)} \approx 0\, ,
\label{momentumconstro}
\end{equation}
therefore we found, again, that the Hamiltonian constraint ${\mathcal H}^{(-3/2)}$ and the momentum constraints ${\mathcal H}_{i}^{(-3/2)}$ are secondary Dirac's constraints.

The next step is the preservation of the primary constraint $C_{\phi}$ defined in \eqref{extraprimary} 
\begin{eqnarray}
\label{vincconf}
{\dot C}_{\phi}&=&\left\{C_{\phi},H_{T}^{(-3/2)}\right\}\\
&=&\left\{C_{\phi},\int d^3 x N^i{\mathcal{H}_{i}^{(-3/2)}}\right\}+\left\{C_{\phi},\int d^3x N{\mathcal{H}^{(-3/2)}}\right\}\nonumber\,,
\end{eqnarray}
having $C_{\phi}$ non-zero Poisson brackets with the Hamiltonian constraint $\mathcal{H}^{(-3/2)}$ and the momentum constraints $\mathcal{H}_{i}^{(-3/2)}$. 
In Appendix A we calculate these Poisson brackets using smearing functions, in particular we employ a generic, non null, function $f(x)$ for $C_{\phi}$ and the shift functions $N^{i}(x)$ for ${\mathcal H}_{i}^{(-3/2)}$. The final results are:
\begin{eqnarray}
&&\left\{\int d^3x f(x) C_{\phi}(x),\int d^3 x' N^i(x'){\mathcal{H}_{i}^{(-3/2)}}(x')\right\}\nonumber\\
&&= \int d^{3}y f(y) D_k \left( N^{k}(\pi_{h}-\phi\pi_{\phi}) \right) \nonumber\\
&&= -\int d^{3}y D_k f(y) N^{k}\left(\pi_{h}-\phi\pi_{\phi}\right)\, \approx 0\,.
\label{derivoprimario2}
\end{eqnarray}
 equivalent to 
 \begin{equation}
\left\{C_{\phi}(x),{\mathcal {H}}_{i}^{(-3/2)}(x')\right\}=-\partial'_{i}\delta(x,x')C_{\phi}(x')\,, 
\label{differntial0}
 \end{equation}
 and 
 \begin{eqnarray}
&&\left\{\int d^{3}x f(x)C_{\phi}(x), \int d^{3}x' N(x'){\mathcal{H}}^{(-3/2)}(x')\right\}\nonumber\\ 
&&=\frac{1}{2}\int d^{3}yN(y)f(y){\mathcal{H}}^{(-3/2)}(y)
\approx 0 \, ,
 \label{endin0}
 \end{eqnarray}
which, pairwise, can be re-written in differential form as 
\begin{equation}
\left\{C_{\phi}(x),{\mathcal {H}}^{(-3/2)}(x')\right\}=\frac{1}{2}{\mathcal{H}}^{(-3/2)}(x)\delta(x,x')\,.
\label{differo}
\end{equation}
 
Finally, the condition for the preservation of the primary constrain $C_{\phi}(x)$ is
\begin{eqnarray}
\label{vincsumm}
&&\left\{\int d^{3}x f(x)C_{\phi}(x),H_{T}^{(-3/2)}\right\}\\
&&=-\int d^{3}y D_k f(y)  \left( N^{k}C_{\phi} \right)
+\frac{1}{2}\int d^{3}yN(y)f(y){\mathcal{H}^{(-3/2)}}(y)\approx 0\,.\nonumber
\end{eqnarray}

\subsection{\label{Hamiltonian Constraint} The constraint algebra of the Poisson brackets of the secondary constraints}
Now we calculate the preservation of the secondary constraints along the dynamic. In doing this we will follow reference \cite{Menotti2017} adapted to our case of the Brans-Dicke theory for $\omega=-\frac{3}{2}$. Repeating the calculations, we have 
\begin{equation}
\left\{ h_{ij}(x), \int d^{3}y N^{l}(y){\mathcal {H}}_{l}^{(-3/2)}(y) \right\}={\cal L}_{\mathbf N} h_{ij}(x)\,,
\label{suh1}
\end{equation}
where ${\cal L}_{\mathbf N}$ is the Lie derivative along the three-dimensional vector field ${\mathbf N}$ defined by the shifts functions $N^l$. Analogously as in \cite{Gionti2021}, but with a longer calculation, one has 
\begin{equation}
\left\{ \pi^{ij}(x), \int d^{3}y N^{l}(y){\mathcal {H}}_{l}^{(-3/2)}(y) \right\}={\cal L}_{\mathbf N} \pi^{ij}(x)\,,
\label{suh2}
\end{equation}
We observe that 
\begin{eqnarray}
&&\left\{\phi(x), \int N^l(y) {\mathcal H}_l(y)^{(-3/2)}d^3 y\right\} \label{sufi3} \\
&&=\frac{\delta}{\delta \pi_{\phi}(x)} \int d^3 y \pi_{\phi}(y)D_i\phi(y)N^i (y) 
=N^i (x) D_i\phi(x)={\cal L}_{\mathbf N} \phi(x)\,, \label{sufi4} \nonumber
\end{eqnarray}
while repeating the same reasoning on the momentum $\pi_{\phi}$ conjugated to $\phi$, we obtain 
\begin{eqnarray}
&&\left\{\pi_{\phi}(x), \int N^l(y){ {\mathcal H}_l^{(-3/2)}}(y)d^3 y\right\} \\
&&=-\frac{\delta}{\delta \phi (x)} \int d^3 y \pi_{\phi}(y)D_i\phi(y)N^i (y) \nonumber
=D_i \left(\pi_{\phi}(x)N^i(x)\right)\, .   
\label{supi}
\end{eqnarray}

The momenta calculated by the Legendre transformation using the Lagrangian ${\mathcal {L}}_{ADM}$ \eqref{eq:Lagrangian2} are densities as well, as it is immediate looking at momenta 
\eqref{pippo1}. Then $\frac{\pi_{\phi}}{\sqrt h}$ is a scalar function. So
\begin{eqnarray}
{\cal L}_{\mathbf N} \pi_{\phi}(x)&=&{\cal L}_{\mathbf N}\left(\sqrt{h}\left(\frac{\pi_{\phi}}{\sqrt h}\right)\right)\nonumber \\
&=&{\cal L}_{\mathbf N}(\sqrt h)\frac{\pi_{\phi}}{\sqrt h}+{\sqrt h}{\cal L}_{\mathbf N}\left(\frac{\pi_{\phi}}{\sqrt h}\right)\nonumber\\
&=&{\pi_{\phi}} D_i N^i+{\sqrt {h}}\partial_{i}\left(\frac{\pi_{\phi}}{\sqrt h}\right)N^i  \\
&=&\pi_{\phi}\frac{1}{\sqrt h}\partial_{i}({\sqrt h}N^i)+\partial_{i}(\pi_{\phi})N^i-\frac{\pi_{\phi}}{2 h}\partial_i (h)N^i\nonumber\\
&=&\pi_{\phi} \partial_i N^i+\partial_i (\pi_{\phi})N^i=D_i\left(\pi_{\phi}(x) N^i(x) \right)\,, \nonumber 
\label{tuttipasspi}
\end{eqnarray}
therefore $\int d^{3}x N^{l}{\mathcal {H}}_{l}$ is the generator of space diffeomorphisms on the three surfaces $\Sigma$ of the canonical variables $(h_{ij}, \phi, \pi^{ij}, \pi_{\phi})$, and any function $F(h_{ij}, \phi, \pi^{ij}, \pi_{\phi})$ of them, in particular of the density functions $\mathcal H$ and ${\mathcal H}_i$. Therefore we have 
\begin{equation}
{\cal L}_{\mathbf N}\sqrt{h}=\sqrt{h}D_l N^l\,,
\end{equation}
and then we have
\begin{eqnarray}
&&{\cal L}_{\mathbf N} {\mathcal H}_i^{(-3/2)}={\sqrt h}{\cal L}_{\mathbf N}\frac {{\mathcal H}_i^{(-3/2)}}{\sqrt h}+ {{\mathcal H}_i^{(-3/2)}}D_l N^l\nonumber\\
&&=N^l \partial_l {\mathcal H}_i^{(-3/2)} + {\mathcal H}_l^{(-3/2)} \partial_i N^l + {\mathcal H}_i^{(-3/2)} \partial_l N^l \;,
\label{sumomentum}
\end{eqnarray}
which entitle us to write 
\begin{eqnarray}
 &&\left\{{\mathcal H}_i^{(-3/2)}, \int N^s(y) {\mathcal H}_s^{(-3/2)}(y)d^3 y\right\}\nonumber\\
 &&=N^l \partial_l {\mathcal H}_i^{(-3/2)} + {\mathcal H}_l^{(-3/2)} \partial_i N^l + {\mathcal H}_i^{(-3/2)} {\partial_l} N^l\;,
 \label{informa}
 \end{eqnarray}
and then the constraint algebra among the momenta constraints
\begin{eqnarray}
  &&\left\{{\mathcal H}_i^{(-3/2)}(x), {\mathcal H}_j^{(-3/2)}(x')\right\} \nonumber\\
  &&= {\mathcal H}_i^{(-3/2)}(x') \partial_j \delta(x,x')- {\mathcal H}_i^{(-3/2)}(x) {\partial_j}' \delta(x,x')\;.  
  \label{commuto}
  \end{eqnarray}
  
As regard the Hamiltonian constraint ${\mathcal H}$, we start from the following 
\begin{equation}
  {\cal L}_{\mathbf N}{\mathcal H}^{(-3/2)}= \left\{{\mathcal H}^{(-3/2)}, \int N^s(y) {\mathcal H}_s^{(-3/2)}(y)d^3 y\right\}\;, 
  \label{liehamilt}
  \end{equation}
and repeating the same reasoning above, we get 
\begin{eqnarray}
 {\cal L}_{\mathbf N} {\mathcal H^{(-3/2)}}&=&{\sqrt h}{\cal L}_{\mathbf N}\frac {\mathcal H^{(-3/2)}}{\sqrt h}+\frac {\mathcal H^{(-3/2)}}{\sqrt h} {\cal L}_{\mathbf N}{\sqrt h} \nonumber\\
 &=&N^l \partial_l {\mathcal H^{(-3/2)}} + {\mathcal H^{(-3/2)}} \partial_i N^i \;, 
 \label{rifa}
  \end{eqnarray}  
finally we can write
\begin{equation}
 \left\{{\mathcal H}^{(-3/2)}(x), {\mathcal H}_i^{(-3/2)}(x')\right\}=-{\mathcal H}^{(-3/2)}(x'){{\partial}'_i}\delta(x,x')\,.
 \label{harifa}
 \end{equation}
 
One of most complicated calculation in the canonical analysis of gravitational theories is the Poisson brackets of the Hamiltonian constraint. These brackets, using the lapse $N(x)$ as smearing function, are usually expressed as 
\begin{equation}
\left\{ \int d^{3}x N(x) {\mathcal{H}}^{(-3/2)}(x), \int d^{3}x N'(x') {\mathcal{H}}^{(-3/2)}(x')\right\}\,,
\label{possyhami}
\end{equation}
where $\mathcal{H}^{(-3/2)}$ is the Hamiltonian constraint in \eqref{hamiltoconst}. As it is discussed in \cite{Menotti2017}, non-zero contribution to these Poisson brackets are given by non-algebraic variation  $\delta h_{ij}$ of the metric function multiplied by algebraic variations $\delta \pi^{ij}$ and, similarly, non-algebraic variation $\delta \phi$ of the field with algebraic variation $\delta \pi_{\phi}$ of its relative momentum. 
Detailed discussion is provided in Appendix B where we find that
the Dirac's constraint algebra generated by the Poisson brackets of the Hamiltonian constraint \eqref{possyhami} is 
\begin{eqnarray}
&&\left\{ \int d^{3}x N(x) {\mathcal{H}}^{(-3/2)}(x), \int d^{3}x N'(x') {\mathcal{H}}^{(-3/2)}(x')\right\} \nonumber \\
&=&\int d^{3}y(ND^{i}N'- N'D^{i}N){\mathcal{H}_i^{(-3/2)}}\nonumber\\
&&+\int d^{3}y(ND^{i}N'- N'D^{i}N)(D_i \log\phi)C_{\phi}\,,
\label{finio}
\end{eqnarray}
it can be re-written in differential form 
\begin{eqnarray}
&&\{{\mathcal{H}}^{(-3/2)}(x),{\mathcal{H}}^{(-3/2)}(x')\}\nonumber\\
&=&{\mathcal {H}}^{(-3/2)}_{i}(x)\partial^{i}\delta(x,x')-{\mathcal H}^{(-3/2)}_{i}(x'){\partial}'^{i}\delta(x,x') \nonumber\\
&&+\left[D^{i}(\log\phi(x))\right]C_\phi(x)\partial_{i}\delta(x,x')\nonumber\\
&&-\left[D^{i}(\log\phi(x'))\right]C_\phi(x'){\partial}'_{i}\delta(x,x').
\label{diffeoreo}
\end{eqnarray}

Notice that the algebra of the Hamiltonian constraint contains a first term proportional to the momentum constraints as in Einstein's geometro-dynamics. This term accounts for the evolution of the three dimensional spatial surfaces in four dimensional space. The extra term, proportional  to the primary first clas constraint $C_{\phi}$, is due to conformal invariance of the theory \cite{Henneaux2010}. 

\subsection{\label{JF-EF}Transformations from the Jordan to the Einstein Frame}

As we have already remarked, the following Weyl (conformal) transformation on the metric tensor 
\begin{equation}
{\widetilde g}_{\mu\nu}=\Big(16\pi G f(\phi)\Big)^{\frac{2}{n-2}}g_{\mu\nu}\;, 
\label{Weyltrans2}
\end{equation}
implies that the ADM metric tensor in the Einstein frame is 
\begin{eqnarray}
\widetilde {g}&=&-({\widetilde N}^{2}-{\widetilde N}_{i}\widetilde{N}^{i})dt \otimes dt 
+\widetilde{N}_{i}(dx^{i} \otimes dt
+dt \otimes dx^{i})\nonumber\\
&&+\widetilde{h}_{ij}dx^{i} \otimes dx^{j}\;\;, 
\label{EFmetricADM2}
\end{eqnarray}
where the expression of ${\widetilde N}, {\widetilde N}_{i}, {\widetilde{h}}_{ij}$ can be found in \eqref{tilderelation0}.  
It is not difficult to see that the action \eqref{BDaction2} in the Einstein frame, once one perform the transformation \eqref{Weyltrans2} for the Brans-Dicke case $f(\phi)=\phi$, becomes 
\begin{eqnarray}
S^{(-3/2)}&=&\frac{1}{16\pi G} \int_{M}dx^4\sqrt{-{\widetilde g}}\left({}^4{\widetilde R}-V(\phi)\right)\nonumber\\
&&+\frac{1}{8\pi G}\int_{\partial M}d^{3}x\sqrt{\widetilde h}{\widetilde K}\,, 
\label{BDEFpart}
\end{eqnarray}
where the potential $V(\phi)$, defined in \eqref{AandV}, becomes simply a constant 
\begin{equation}
    {\widetilde {V}}\equiv\frac{\alpha}{16\pi G}\,,
    \label{ricost}
\end{equation}
$\widetilde K$ is the trace of the extrinsic curvature ${\widetilde K}_{ij}$ in the Einstein frame. As usual, we continue to make the $3+1$ decomposition: following the procedure as in Eq. \eqref{scompi}  and the curvature $3+1$ splitting \cite{DeWitt1967} \cite{Esposito1992} the action \eqref{BDEFpart} becomes
\begin{eqnarray}
 S^{(-3/2)}&=&\frac{1}{16\pi G} \int_{R\times \Sigma}dt d^{3}x\sqrt{{\widetilde h}}\widetilde{N}\Big({}^3{\widetilde R} +{\widetilde{K}}^{ij}{\widetilde{K}}_{ij}-{\widetilde{K}}^{2} \nonumber\\ 
 &&-{\widetilde V}\Big)+
 \frac{1}{8\pi G}\int_{\partial M}d^{3}x\sqrt{\widetilde h}{\widetilde K}\,.  
\label{azioneADM}
\end{eqnarray}
The extrinsic curvature ${\widetilde K}_{ij}$ is, of course, defined as follows 
\begin{equation}
{\widetilde K}_{ij}=\frac{1}{2 \widetilde{N}}
\left(-\frac{\partial \widetilde{h}_{ij}}{\partial t} +{\widetilde{D}}_{i}{\widetilde{N}}_{j} +{\widetilde{D}}_{j}{\widetilde{N}}_{i}\right),
\label{conformalextrins}
\end{equation}
where ${\widetilde{D}}$ is the covariant derivative defined with the Levi-Civita connection ${\widetilde{\Gamma}}^{c}_{ab}$ corresponding to the metric tensor ${\widetilde{h}}_{ij}$. Following \cite{Dabrowski2008}, the relation between ${\widetilde{\Gamma}}^{c}_{ab}$ and ${\Gamma}^{c}_{ab}$ is 
\begin{eqnarray}
{\widetilde{\Gamma}}^{c}_{ab}&=&{\Gamma}^{c}_{ab}+\frac{1}{(16\pi G \phi)^{\frac{1}{2}}}
\Big(\delta^{c}_{a}[(16\pi G \phi)^{\frac{1}{2}}]_{,b} + 
\delta^{c}_{b}[(16\pi G \phi)^{\frac{1}{2}}]_{,a} \nonumber \\
&& - h_{ab}h^{cd}[(16\pi G \phi)^{\frac{1}{2}}]_{,d}\Big).
\label{conneconformo}
\end{eqnarray}
Using last equation and relations \eqref{tilderelation0}, one easily finds 
\begin{equation}
{\widetilde{K}}_{ij}=(16\pi G \phi)^{\frac{1}{2}}K_{ij}-\frac{16\pi G\, h_{ij}}{2N (16\pi G \phi)^{\frac{1}{2}}}\left(\dot{\phi}-N^{l}\partial_{l}\phi\right).
\label{estrinsconform}
\end{equation}
The action \eqref{azioneADM} indicated that the ADM-Lagrangian density function ${\widetilde{\mathcal{L}}}_{ADM}^{(-3/2)}$ in the Einstein frame is
\begin{equation}
{\widetilde{\mathcal{L}}}_{ADM}^{(-3/2)}=\frac{1}{16\pi G}\sqrt{{\widetilde h}}\widetilde{N}\left({}^3{\widetilde R} +{\widetilde{K}}^{ij}{\widetilde{K}}_{ij}-{\widetilde{K}}^{2} -{\widetilde V}\right)\,,
\label{ADMdensityEF}
\end{equation}
The definition of the momenta ${\widetilde{\pi}}^{ij}$ associated to ${\widetilde{h}}_{ij}$ is 
\begin{equation}
{\widetilde \pi}^{ij}= \frac{\partial {\mathcal {\widetilde L}}_{ADM}^{(-3/2)} }{\partial \dot{\widetilde{h}}_{ij}}=
-\frac{\sqrt{\widetilde {h}}}{{16 \pi G}}\left( {\widetilde K}^{ij}-{\widetilde K}{\widetilde h}^{ij}\right),
\label{momo}
\end{equation}
which using \eqref{estrinsconform} and definition \eqref{pippo1}, implies that
\begin{equation}
{\widetilde \pi}^{ij}=\frac{1}{16\pi G \phi}\pi^{ij} .
\label{regoltrans}
\end{equation}

In the case $\omega \neq -\frac{3}{2}$ \cite{Gionti2021} the momentum ${\widetilde\pi}_{\phi}$ conjugated to $\phi$ is defined as
\begin{eqnarray}
{\widetilde \pi}_\phi&=&\frac{\partial {\mathcal {\widetilde L}}_{ADM}}{\partial \dot{\phi}}=\frac{\sqrt{\widetilde {h}}(\omega +\frac{3}{2})}{8\pi G {\widetilde N}{\phi}^2}\left(\dot{\phi}-{\widetilde N}^i\partial_i\phi \right)\nonumber \\
&=&\frac{1}{\phi}(\phi \pi_{\phi}-\pi_{h})\,, 
\label{transforno}
\end{eqnarray}
therefore it is evident that $-\frac{C_{\phi}}{\phi}$ is mapped into ${\widetilde \pi}_\phi$ by the Weyl (conformal) transformation. Then in the particular case $\omega = -\frac{3}{2}$, the primary constraint in the Einstein frame
\begin{equation}
{\widetilde \pi}_\phi=\frac{\partial {\mathcal {\widetilde L}}_{ADM}^{(-3/2)}}{\partial \dot{\phi}}\approx 0 \,,
\label{miaccorgo}
\end{equation}
corresponds to $C_\phi\approx 0$ in the Jordan frame. The canonical Hamiltonian density $\widetilde{{\mathcal{H}}}_c$ \cite{dirac1966} \cite{Esposito1992} coincides with the ADM Hamiltonian density $\widetilde{{\mathcal{H}}}_{ADM}$, in analogy to the corresponding definition in the Jordan frame \eqref{hamiltodefin}, and it is defined on the constraint surface where the primary constraints are zero
\begin{equation}
\widetilde{\pi}_N =\frac{\partial \widetilde{{\mathcal {L}}}_{ADM}^{(-3/2)}}{\partial \dot{\widetilde{N}}}\approx 0 \ , {\widetilde{\pi}}_i=\frac{\partial \widetilde{{\mathcal L}}_{ADM}^{(-3/2)}}{\partial \dot{\widetilde {N}}^i}\approx 0 \, , \widetilde{\pi}_{\phi} \approx 0\,,
\label{primarysurfaceEF}
\end{equation}
so we have 
\begin{eqnarray}
{\widetilde{\mathcal{H}}}_{ADM}^{(-3/2)}&=&\widetilde{{\pi}}^{ij}{\dot {\widetilde{{h}}}}_{ij}-\widetilde{\mathcal{L}}_{ADM}^{(-3/2)}\nonumber\\
&=&\frac{\sqrt{{\widetilde h}}\widetilde{N}}{16\pi G}\Bigg[ -{}^{3}{\widetilde R}+\frac{(16\pi G)^2}{\widetilde h}\left( {\widetilde \pi}^{ij}{\widetilde \pi}_{ij}-\frac{{{\widetilde \pi}_h}^2}{2}\right) 
+{\widetilde V}\Bigg] \nonumber\\
 &&-2\widetilde{N}^i{\widetilde D}_j{\widetilde \pi}^{j}_{i}\nonumber\\
 &=&{\widetilde{N}}{\widetilde{\mathcal H}}^{(-3/2)}+{\widetilde{N}}^{i}\widetilde{{\mathcal H}}_{i}^{(-3/2)}\,,
\label{HamiltodensityBDEF}
\end{eqnarray}
where now the Hamiltonian constraint ${\widetilde{\mathcal H}}^{(-3/2)}$ is 
\begin{equation}
{\widetilde{\mathcal H}}^{(-3/2)}=\frac{\sqrt{{\widetilde h}}}{16\pi G}\Big( -{}^{3}{\widetilde R}+\frac{(16\pi G)^2}{\widetilde h}\big( {\widetilde \pi}^{ij}{\widetilde \pi}_{ij}-\frac{{{\widetilde \pi}_h}^2}{2}\big) 
+{\widetilde V}\Big)\,, 
\label{hamiltoEF}
\end{equation}
while the momentum constraints $\widetilde{{\mathcal H}}_{i}^{(-3/2)}$ are
\begin{equation}
 \widetilde{{\mathcal H}}_{i}^{(-3/2)}=-2{\widetilde D}_j{\widetilde \pi}^{j}_{i} \,.
\label{momentaEF}
\end{equation}

The ADM-Hamiltonian density ${\widetilde{\mathcal{H}}}_{ADM}^{(-3/2)}$ has the same functional form as ${\mathcal{H}}_{ADM}$ in Einstein Geometro-dynamics \cite{Menotti2017} plus a constant term ${\widetilde V}$ which looks as a cosmological constant in \eqref{hamiltoEF}. 
The total Hamiltonian density ${\widetilde H}_T^{(-3/2)}$ is a linear combination of the canonical Hamiltonian with the primary constraints with Lagrange multipliers \cite{dirac1966} \cite{Esposito1992}
\begin{eqnarray}
{\widetilde H}_T^{(-3/2)}&=&\int d^{3}x\Big({\widetilde{\lambda}}{\widetilde \pi}_N+{\widetilde \lambda}^{i} {\widetilde \pi} _{i}+{\widetilde{\lambda}}_{\phi}\widetilde{C}_{\phi}\nonumber\\
&&+
{\widetilde{N}}{\widetilde{{\mathcal {H}}}}^{(-3/2)}+{\widetilde{N}}^{i}{\widetilde{{\mathcal {H}}}}_{i}^{(-3/2)}\Big),
\label{hamiltonianatotaleEF}
\end{eqnarray}
confronting and contrasting it with the analogous quantity in the Jordan frame \eqref{hamiltonianatot1},  we, again, notice that $\widetilde{C}_{\phi}\equiv-\widetilde{\phi}\widetilde{\pi}_\phi$ through Weyl (conformal) Hamiltonian transformations. It is very easy to see, in analogy to the equations \eqref{hamiltoconst} and \eqref{momentumconstro}, the Hamiltonian constraint $\widetilde{\mathcal{H}}$ and the momentum constraints ${\widetilde{\mathcal{H}}}_i$ are secondary constraints; quite straightforward to see, as well, is the fact that 
\begin{equation}
\dot{\widetilde{C}}_{\phi}=\left\{\widetilde{C}_{\phi},{\widetilde H}_T^{(-3/2)}\right\}=0\,.
\label{primaryconserv}
\end{equation}

Previous observations, cfr. \eqref{HamiltodensityBDEF}, has remarked the equivalence of this Brans-Dicke theory with $\omega=-\frac{3}{2}$ in the Einstein frame with Einstein General Relativity. Based on this equivalence, it is clear the Poisson brackets constraint algebra among secondary first class constraints is 
\begin{eqnarray}
&&\left\{\widetilde{C}_{\phi},{\widetilde{\mathcal{H}}}^{(-3/2)}\right\}=0\;,
\left\{\widetilde{C}_{\phi},{\widetilde{\mathcal{H}}}_i^{(-3/2)}\right\}=0\,,\nonumber \\
&&\{\widetilde{{\cal{H}}}_{i}^{(-3/2)}(x), {\widetilde{\cal{H}}}_{j}^{(-3/2)}(x')\} \nonumber\\
&&={\widetilde{\cal{H}}}_{i}^{(-3/2)}(x')\partial_j\delta(x,x')
-{\widetilde{\cal{H}}}_{j}^{(-3/2)}(x)\partial_i\delta(x,x')\,,\label{algebraconstr}  \\
&&\{{\widetilde{\cal{H}}}^{(-3/2)}(x), {\widetilde{\cal{H}}}_{j}^{(-3/2)}(x')\}=
-{\widetilde{\cal{H}}}^{(-3/2)}(x')\partial'_j\delta(x',x)\,,\nonumber \\
&&\{{\widetilde{\cal{H}}}^{(-3/2)}(x),{\widetilde{\cal{H}}}^{(-3/2)}(x')\}\nonumber\\
&&={\widetilde{\cal H}}^{(-3/2)i}(x)\partial_{i}\delta(x,x')-{\widetilde{\cal H}^{(-3/2)i}}(x')\partial'_{i}\delta(x,x')\,.\nonumber 
\end{eqnarray}
Note the first two Poisson brackets are identically zero and not weakly zero. 

\subsection{\label{canonical2}Canonical transformations and the passage from the  Jordan to the  Einstein Frame}

The transformation formulae from the Jordan to the Einstein frame in the canonical (Hamiltonian) formalism - 
see Eqs. \eqref{tilderelation0}, \eqref{regoltrans}, and \eqref{primarysurfaceEF} -
are 
\begin{eqnarray}
&&\widetilde{N}=N(16\pi G\phi)^{\frac{1}{2}}\,,\;
\widetilde{N}_i=N_i(16\pi G \phi)\,,\nonumber\\
&&{\widetilde{h}}_{ij}=\big(16\pi G \phi \big)h_{ij}\,,\;{\widetilde{\pi}}_N=\frac{\pi_N}{(16\pi G\phi)^{\frac{1}{2}}}\,,\label{trasformoEF}\\
&&{\widetilde{\pi}}_{i}=\frac{{\pi}_{i}}{16\pi G\phi}\,,\;{\widetilde \pi}^{ij}=\frac{{\pi}^{ij}}{16\pi G\phi}\,,\; \widetilde{\phi}=\phi\,,\widetilde{\pi}_{\phi} \approx 0\, . \nonumber
\end{eqnarray}
In the Hamiltonian theory, a transformation $(Q^i(q,p), P_i(q,p))$ between two sets of variables $(q^i,p_i)$ and $(Q^i,P_i)$  is canonical if the \textquotedblleft symplectic two form\textquotedblright
$\omega=dq^i \wedge dp_i$ is invariant -
that is $\omega=dQ^i \wedge dP_i$ - which is equivalent to say that the Poisson brackets fulfill the following conditions
\begin{eqnarray}
\{Q^{i}(q,p),P_{j}(q,p)\}_{q,p}&=&\delta^{i}_{j}\,, \label{symplectic}  \\
\{Q^{i}(q,p),Q^{j}(q,p)\}_{q,p}&=&\{P_{i}(q,p),P_{j}(q,p)\}_{q,p}=0\, .
\nonumber
\label{canonicalone}
\end{eqnarray}
These properties of the canonical transformations imply that if one computes the Poisson brackets between two functions respect to two different sets of variables connected by a canonical transformation,the final results will be the same.

First we notice that the Poisson brackets between the conformal constraint $C_\phi$ and the momentum constraints ${\mathcal{H}_{i}^{(-3/2)}}$ in the Jordan frame are 
\begin{equation}
\left\{C_{\phi}(x),{\mathcal {H}}_{i}^{(-3/2)}(x')\right\}=-\partial'_{i}\delta(x,x')C_{\phi}(x')\,,
\label{ricalco1}
\end{equation}
while the analogous Poisson brackets in the Einstein frame (recall here $\widetilde{C}_{\phi}\equiv-\widetilde{\phi}\widetilde{\pi}_\phi$)
\begin{equation}
\left\{{\widetilde C}_{\phi}(x),{\widetilde{{\mathcal {H}}}_{i}}^{(-3/2)}(x')\right\}=0\,.
\label{0EFi}
\end{equation}

In analogous way, the Poisson brackets of the conformal constraint $C_\phi$ with the Hamiltonian constraint ${\mathcal{H}^{(-3/2)}} $ is 
\begin{equation}
\left\{C_{\phi}(x),{\mathcal {H}}^{(-3/2)}(x')\right\}=\frac{1}{2}{\mathcal{H}}^{(-3/2)}(x)\delta(x,x')\,,
\label{ricalco2}
\end{equation}
the same Poisson bracket in the Einstein frame is 
\begin{equation}
\left\{\widetilde{C}_{\phi}(x),{\widetilde{{\mathcal {H}}}}^{(-3/2)}(x')\right\}=0\,.
\label{0EF}
\end{equation}

The last couple of non-equivalent Poisson brackets in the two frames are the Poisson brackets of two Hamiltonian constraints ${\mathcal {H}}^{(-3/2)}$ evaluated in two different points. In the Jordan frame, we have 
\begin{eqnarray}
&&\{{\mathcal{H}}^{(-3/2)}(x),{\mathcal{H}}^{(-3/2)}(x')\}\nonumber\\
&=&{\mathcal {H}}^{(-3/2)}_{i}(x)\partial^{i}\delta(x,x')-{\mathcal H}^{(-3/2)}_{i}(x'){\partial}'^{i}\delta(x,x') \nonumber\\
&&+\left[D^{i}(\log\phi(x))\right]C_\phi(x)\partial_{i}\delta(x,x')\nonumber\\
&&-\left[D^{i}(\log\phi(x'))\right]C_\phi(x'){\partial}'_{i}\delta(x,x')\,, 
\label{diffeoreo1}
\end{eqnarray}
and in the Einstein frame 
\begin{eqnarray}
&&\{{\widetilde{\cal{H}}}^{(-3/2)}(x),{\widetilde{\cal{H}}}^{(-3/2)}(x')\}\nonumber\\
&&={\widetilde{\cal H}}^{(-3/2)i}(x)\partial_{i}\delta(x,x')-{\widetilde{\cal H}^{(-3/2)i}}(x')\partial'_{i}\delta(x,x')\,.
\label{duos}
\end{eqnarray}

This in-equivalence of the three Poisson brackets in the two frames is a clear evidence that the Hamiltonian transformations \eqref{trasformoEF} are not - strictly speaking - a set of canonical transformations.  

This result could appear a bit awkward. In fact in the section \ref{Conformal21}, we showed that the Brans-Dicke action in the case $\omega=-\frac{3}{2}$ \eqref{BDaction} is invariant under Weyl-(conformal) transformations on the metric $g_{\mu\nu}$ \eqref{conformal} and on the scalar filed $\phi$ \eqref{campoconfo}.
The transformations from the Jordan to the Einstein frame, given by \eqref{Weyltrans2} on the metric tensor and holding the scalar field $\widetilde{\phi}=\phi$, corresponds to the particular Weyl (conformal) transformations \eqref{conformal} and  \eqref{campoconfo}
\begin{equation}
\Omega=(16\pi G \phi)^{\frac{1}{2}}\,,\; \widetilde{\phi}=\frac{1}{16\pi G}\,; 
\label{specilaconf}
\end{equation}
when one passes from the action \eqref{BDaction} to \eqref{finale}. This conformal symmetry should map solutions of the equations of motion  into solutions of the equations of motion. In other words, the conformal symmetry should be an automorphism among solutions of the equations of motion. Therefore, the transformations from the Jordan to the Einstein frames should be canonical transformations. Since $\widetilde{\phi}=\frac{1}{16\pi G}$, the conformal transformations generate a set of new Hamiltonian variables $(Q^i(q,p), P_i(q,p))$ such that     
\begin{equation}
\det \left\lvert \frac{\partial(Q^i(q,p), P_i(q,p)}{\partial (q,p)} \right\rvert=0.
\label{sing}
\end{equation}
This implies that these transformations are  singular and the automorphism does not hold in this case. 

\section{Summary and Conclusions}
\label{Conclusions}
This essay aimed to continue, with further details, a project started in \cite{Gionti2021}. The central point has been to scrutinize whether the transformations from the Jordan to the Einstein frames are Hamiltonian canonical transformations as claimed in 
\cite{Olmo}\cite{Deruelle2009}$ $\cite{Deruelle2010}. 
The tool, we employed, was the Hamiltonian Dirac's constraint analysis of the Brans-Dicke theory in the particular case $\omega =-\frac{3}{2}$. In order to fulfill the goals of our project, we started with a brief summary of the results of the Hamiltonian Dirac's constraint analysis of the Brans-Dicke theory in the case $\omega \neq -\frac{3}{2}$ \cite{Gionti2021}. We showed, as it is extensively argued in \cite{Kuchar1} \cite{Kuchar2}, that the algebra of the Poisson brackets of the secondary first class constraints is the same as Einstein's geometro-dynamics \eqref{differential}. This could suggest that the transformations from the Jordan to the Einstein frames are canonical transformations in the Hamiltonian theory \cite{Deruelle2009}. Unfortunately, this belief is wrong \eqref{noncanonicalcond}. Therefore, one cannot perform the Hamiltonian Dirac's constraint analysis in the Einstein frame, where the calculations are simpler, and pretend that it is the same in the Jordan frame \cite{Garay1992}. A set of Hamiltonian canonical transformations does exist. They are called the Anti-Newtonian gravity transformations \eqref{tredimesconf} \cite{Niedermaier2019}\cite{Niedermaier2020} and differ from the conformal transformations \eqref{tilderelation0}. This in-equivalence of the Jordan and Einstein frames for Brans-Dicke theory in the Hamiltonian formalism addresses quantum in-equivalence \cite{Falls2018} \cite{Kamenshchik2014} \cite{Filippo2013} \cite{Banerjee2016} as well although, even at quantum level, there are several papers arguing in favor of quantum equivalence \cite{Ohta2017} \cite{ChristianS2017}\cite{Ruf2017}. 

\begin{widetext}
\begin{table}[ht]
\centering
\begin{tabular}{|c|c|}
 \hline 
  \multicolumn{2}{|c|}{\bf Hamiltonian Analysis of BD for $\omega \neq -\frac{3}{2}$} \\
  {{\bf in Jordan Frame} }    &  {{\bf  in Einstein Frame} }  \\
  \hline
  {{\it constraints}}      & {{\it constraints}}   \\
  {$\pi_N \approx 0; \pi^{i}\approx 0; \mathcal{H}\approx 0;\mathcal{H}_i\approx 0; $}   &  
  {${\widetilde{\pi}_N} \approx 0; {\widetilde{\pi}}_{i}\approx 0; \widetilde{\mathcal{H}}\approx 0;{\widetilde{\mathcal{H}}}_i\approx 0;$}   \\
  \hline
  {{\it constraint algebra }}     &  {{\it constraint algebra}}   \\
    $\{\pi_N,\pi_{i}\}= 0; \{\pi_N,{\mathcal{H}}\}= 0; \{\pi_N,{\mathcal{H}}_i\}= 0; \{\pi_{i},\mathcal{H}\}= 0;$  
  & $\{\widetilde{\pi}_N,\widetilde{\pi_{i}}\}= 0; \{\widetilde{\pi}_N,\widetilde{{\mathcal{H}}}\}= 0; \{\widetilde{\pi}_N,\widetilde{{\mathcal{H}}}_i\}= 0; \{{\widetilde{\pi}}_{i},\widetilde{\mathcal{H}}\}= 0;$ \\
  $\{{\pi}_i,{\mathcal{H}}_j\}= 0;\,\left\{{\mathcal H}(x), {\mathcal H}_i(x')\right\}=-{\mathcal H}(x'){{\partial}'_i}\delta(x,x');$ & 
  $\{{\widetilde{{\pi}}}_i,{\widetilde{{\mathcal{H}}}}_j\}= 0;\,\left\{{\widetilde{{\mathcal H}}}(x), 
  {\widetilde{{\mathcal H}}}_i(x')\right\}=
  -{\widetilde{{\mathcal{H}}}}(x'){{\partial}'_i}\delta(x,x');$ \\
  $\left\{{\mathcal H}_i(x), {\mathcal H}_j(x')\right\} ={\mathcal H}_i(x') \partial_j \delta(x,x')- {\mathcal H}_j(x) {\partial_i}' \delta(x,x');$ & 
  $\{{\widetilde{{\mathcal H}}}_i(x), {\widetilde{{\mathcal H}}}_j(x')\} ={\widetilde{{\mathcal H}}}_i(x') \partial_j \delta(x,x')- {\widetilde{{\mathcal H}}}_i(x) {\partial_i}' \delta(x,x');$\\
  $\{{\mathcal{H}}(x),{\mathcal{H}}(x')\}={\mathcal {H}}^{i}(x)\partial_{i}\delta(x,x')-{\mathcal H}^{i}(x'){\partial}'_{i}\delta(x,x') ;$ & $\{{\widetilde{{\mathcal{H}}}}(x),{\widetilde{{\mathcal{H}}}}(x')\}
  ={\widetilde{{\mathcal H}}}^i(x)\partial_{i}\delta(x,x')-{\widetilde{{\mathcal H}}}^i(x'){\partial}'_{i}\delta(x,x'); $\\
  \hline 
\end{tabular}
\caption{Dirac's constraints and constraint algebra in Jordan and Einstein frames for $\omega\neq -\frac{3}{2}$ (see ref \cite{Gionti2021} for details).}
\label{tab:summary_1}
\end{table}
    
\begin{table}[ht]
  \centering
\begin{tabular}{|c|c|}
\hline
\multicolumn{2}{|c|}{\bf Hamiltonian Analysis of BD for $\omega = -\frac{3}{2}$}\\
 {{\bf in Jordan Frame} }    &  {{\bf in Einstein Frame} }  \\
  \hline
  {{\it constraints}}      & {{\it constraints}}   \\
  {$\pi_N \approx 0; \pi^{i}\approx 0; C_{\phi}\approx 0; \mathcal{H}^{(-3/2)}\approx 0;\mathcal{H}^{(-3/2)}_i\approx 0; $}   &  
  {${\widetilde{\pi}_N} \approx 0; {\widetilde{\pi}}_{i}\approx 0;{\widetilde{C}}_{\phi}=-\widetilde{\phi}{\widetilde{\pi}}_{\phi}\approx 0; \widetilde{\mathcal{H}}^{(-3/2)}\approx 0;{\widetilde{\mathcal{H}}^{(-3/2)}}_i\approx 0;$}   \\
  \hline
  {{\it constraint algebra }}     &  {{\it constraint algebra}}   \\
  $\{\pi_N,\pi_{i}\}=\{\pi_N,{\mathcal{H}}^{(-3/2)}\}=\{\pi_N,{\mathcal{H}}^{(-3/2)}_i\}=0;$  
  & $\{\widetilde{\pi}_N,\widetilde{\pi_{i}}\}=\{\widetilde{\pi}_N,\widetilde{\mathcal{H}}^{(-3/2)}\}= 0; \{\widetilde{\pi}_N,\widetilde{{\mathcal{H}}}_i^{(-3/2)}\}= 0;$ \\
  $\{\pi_{i},\mathcal{H}^{(-3/2)}\}=\{{\pi}_i,{\mathcal{H}}^{(-3/2)}_j\}=0;$
  &$\{{\widetilde{\pi}}_{i},{\widetilde{\mathcal{H}}}^{(-3/2)}\}=\{{\widetilde{{\pi}}}_i,{\widetilde{{\mathcal{H}}}}^{(-3/2)}_{j}\}=0;$\\
  $\left\{C_{\phi}(x),{\mathcal {H}}_{i}^{(-3/2)}(x')\right\}=-\partial'_{i}\delta(x,x')C_{\phi}(x');$ &$\left\{{\widetilde C}_{\phi}(x),{\widetilde{{\mathcal {H}}}_{i}}^{(-3/2)}(x')\right\}=0;$\\
  $ \left\{C_{\phi}(x),{\mathcal {H}}^{(-3/2)}(x')\right\}=\frac{1}{2}{\mathcal{H}}^{(-3/2)}(x)\delta(x,x');$ & 
   $\left\{\widetilde{C}_{\phi}(x),{\widetilde{{\mathcal {H}}}}^{(-3/2)}(x')\right\}=0;$ \\
  $\left\{{\mathcal H}^{(-3/2)}(x), {\mathcal H}^{(-3/2)}_i(x')\right\}=-{\mathcal H}^{(-3/2)}(x'){{\partial}'_i}\delta(x,x');$ & 
  $\left\{{\widetilde{{\mathcal H}}}^{(-3/2)}(x), 
  {\widetilde{{\mathcal H}}}^{(-3/2)}_i(x')\right\}=
  -{\widetilde{{\mathcal{H}}}}^{(-3/2)}(x'){{\partial}'_i}\delta(x,x');$ \\
  $\left\{{\mathcal H}^{(-3/2)}_i(x), {\mathcal H}^{(-3/2)}_j(x')\right\} ={\mathcal H}^{(-3/2)}_i(x') \partial_j \delta(x,x')$ & 
  $\{{\widetilde{{\mathcal H}}}^{(-3/2)}_i(x), {\widetilde{{\mathcal H}}}^{(-3/2)}_j(x')\} ={\widetilde{{\mathcal H}}}^{(-3/2)}_i(x') \partial_j \delta(x,x')$\\
  $ - {\mathcal H}^{(-3/2)}_j(x) {\partial_i}' \delta(x,x');$ & $- {\widetilde{{\mathcal H}}}^{(-3/2)}_i(x) {\partial_i}' \delta(x,x'); $\\
  $\{{\mathcal{H}}^{(-3/2)}(x),{\mathcal{H}}^{(-3/2)}(x')\}=$ & $\{{\widetilde{{\mathcal{H}}}}^{(-3/2)}(x),{\widetilde{{\mathcal{H}}}}^{(-3/2)}(x')\}
  = $\\
  ${\mathcal {H}}^{(-3/2)}_{i}(x)\partial^{i}\delta(x,x')-{\mathcal H}^{(-3/2)}_{i}(x'){\partial}'^{i}\delta(x,x')+  $ & ${\widetilde{{\mathcal H}}}^{(-3/2)}_i(x)\partial^{i}\delta(x,x')-{\widetilde{{\mathcal H}}}^{(-3/2)}_i(x'){\partial}'_{i}\delta(x,x'); $\\
  $\left[D^{i}(\log\phi(x))\right]C_\phi(x)\partial_{i}\delta(x,x')$ & ${ }$\\
  $-\left[D^{i}(\log\phi(x'))\right]C_\phi(x'){\partial}'_{i}\delta(x,x'); $& ${ }$\\
  \hline 
\end{tabular}
\caption{Dirac's constraints and constraint algebra in Jordan and Einstein frames for $\omega= -\frac{3}{2}$}
\label{tab:summary_2}
\end{table}
\end{widetext}

In literature \cite{Gielen} \cite{thiemann2007}, it is claimed that the lapse $N$ and the shifts $N^i$ functions behave as ``gauge'' variables. They appear only as Lagrangian multipliers in the total Hamiltonian \eqref{hamiltonianatot1} after having performed Dirac's constraint's analysis. Therefore  there is a general tendency  to discard these variables and their respective momenta $\pi_N$ and $\pi_{i}$ as unessential. This is the main reason for which in \cite{Deruelle2009} the calculation to verify the conditions for canonical transformations has been performed excluding these variables and arriving to the conclusion that the Hamiltonian  transformations from the Jordan to the Einstein frames are canonical. 
In our opinion, there is a fallacy of this line of thought.
First of all, if one discards the lapse $N$ and the shift $N^i$,
how could one pin down the ADM metric tensor coefficients in \eqref{ADMmetricJF} since they are functions of the lapse and the shifts? In other words, we need to fix the values of the lapse and the shifts, perform a gauge fixing, in order to determine the values of the metric coefficients.
Secondly, the lapse and the shifts act as true canonical variables since by preserving the momenta conjugated to the lapse and the shifts one derives the Hamiltonian and the momentum constraints. 
Maybe, these authors have in mind an example employed by Dirac in his lecture notes \cite{dirac1966}. He considered a couple of second class constraint $q^1 \approx 0$ and $p_1 \approx 0$ and he got rid of them by simply imposing them strongly and eliminating them as variables. He himself stressed that it works only for this particular couple of canonical variables, but not, necessarily, in general. Moreover, the lapse and the shifts are not constraints.

The general impression is that these kind of reasoning presuppose there exists a transformation which divides the set of variables into gauge variables and true (physical) degrees of freedom of the dynamical system. In \cite{Lusanna2017} \cite{Shanmugadhasan1973} \cite{Faddeev:1980be}, it has been extensively shown that there exist local canonical transformations which map the first class constraints into momenta variables $P_a\approx 0$ and the second class constraints into position and momenta variables $q^i\approx 0\,p_i\approx 0$. In \cite{Faddeev:1980be}, it is shown that, in order to find the true, physical, degrees of freedom, one has to make a gauge fixing for each first class constraint (see also \cite{Bonanno2017}). After treating the gauge fixings as a secondary constraints, the Poisson brackets between each previous first class constraint and its relative gauge fixing is non-zero. Then it is possible to find, locally, a set of variables $q^{\star}\,p_{\star}$ which are the true physical degrees of freedom of the system. Therefore, there exist  local transformations capable of dividing the set of canonical variables into two disjoints sets: the physical variables and the gauge variables. Finally, one defines Dirac's brackets using all constraints, which now are second class constraints, and substitutes Poisson brackets with Dirac brackets \cite{dirac1966}. At this point, one can solve ``strongly'' the second class constraints (use each of them to define one variable as function of the others) and reduce the degrees of freedom of the system without any danger of generating inconsistencies. 

The main argument of this essay was to study Jordan and Einstein frames under the light of the  canonical analysis of the Brans-Dicke theory in the case $\omega=-\frac{3}{2}$ with Gibbons-Hawking-York boundary term. We showed explicitly that this theory has a Weyl-(conformal) symmetry. The Dirac's constraint analysis was done and we found five primary constraints and four secondary constraints. All Dirac's constraints are first class constraints according to Dirac's classification. The extra first class constraint is generated by the extra Weyl symmetry. This study was carried out through functional definition of the Poisson brackets following \cite{Menotti2017} and gave several results in agreement to \cite{Gielen}, as regard the coefficients of the Dirac's constraint algebra, but also some differences, the main one being the Poisson bracket \eqref{differo}. On the contrary, we seems to be in perfect agreement with \cite{Zhang2011}. 

The $\omega=-\frac{3}{2}$ Brans-Dicke action \eqref{BDaction}, transformed through the conformal map \eqref{Weyltrans1}  from the Jordan to the Einstein Frame, becomes Einstein's theory of General Relativity with a constant potential that acts as a cosmological constant. The main argument for which Jordan and Einstein frames are not canonically equivalent for $\omega\neq\frac{3}{2}$ holds also in this case. Furthermore, in the Einstein frame, the algebra of the Dirac's constraints \eqref{algebraconstr} is different respect to the Jordan frame \eqref{finio} \eqref{harifa} \eqref{commuto}. This is even a more crystal clear evidence of the non-canonicity.  This may sound strange since it was shown that $\omega=-\frac{3}{2}$ Branse-Dicke is conformal invariant. The puzzle is solved once one notice the conformal mapping \eqref{conformal} \eqref{campoconfo} transforms also the scalar field $\phi(x)$, while the map \eqref{Weyltrans1} keeps $\phi$ unchanged. The conformal transformations form the Jordan to the Einstein frames are a particular case of the conformal transformations \eqref{conformal} \eqref{campoconfo} for $\Omega$ given by \eqref{specilaconf}. However, these transformations  are singular \eqref{sing} for this value of $\Omega$, and then it is not guaranteed this map preserves the structure of the Poisson brackets.\\
A further application of this research could be the analysis to a FLRW mini-supersapce model of the Brans-Dicke theory. Mainly, it is interesting for physical reasons. This analysis could probe several compelling questions. In fact, one could ask: in which sense the Hamiltonian transformations form the Jordan to the Einstein frames are not canonical? We have already seen \cite{Gionti2021} that, in the case $\omega\neq-3/2$, the canonical structure of Poisson brackets is not preserved. That would it happen in the case $\omega=-3/2$ and what could we say regarding the equations of motion? 
Moreover, in this simpler finite dimensional case, we could separate more clearly physical degrees of freedom from gauge degree of freedom by making a guage fixing on the lapse function and defining Dirac's brackets. One could check if the Hamiltonian transformations from the Jordan to the Einstein frames could be canonical and what this physically would mean in light of all previous considerations. Finally, what implication does this Hamiltonian analysis have for the corresponding Lagrangian theory?

\begin{acknowledgments}
We thank Alfio Bonanno and Alexander Kamenshchik for useful discussions.
\end{acknowledgments}
\section*{Appendix}

\subsection{Preservation of the conformal constraint $C_\phi$}
\label{AppA}
Let's begin to calculate the Poisson bracket $\{C_{\phi},N^i{\mathcal{H}_{i}}^{(-3/2)}\}$, first term of Eq. \eqref{vincconf}: 

\begin{widetext}
\begin{eqnarray}
&&\left\{\int d^3x f(x) C_{\phi}(x),\int d^3 x' N^i(x'){\mathcal{H}_{i}^{(-3/2)}}(x')\right\}\nonumber\\
&=&\left\{\int d^{3}x f(x)\left({h}_{ij}{\pi}^{ij}-\phi\pi_{\phi}\right),\int d^{3}{x'}N^{s}(x')\left(-2h_{sb}(x')D_a 
\pi^{ab}(x')+\partial_s\phi \pi_{\phi}\right) \right\}\nonumber\\
&=&\left\{\int d^{3}x f(x)\left({h}_{ij}{\pi}^{ij}\right),\int d^{3}{x'}N^{s}(x')\left(-2h_{sb}(x')D_a \pi^{ab}(x')\right)\right\}\nonumber\\
&& -\left\{\int d^{3}x f(x)\left(\phi\pi_{\phi}\right),\int d^{3}{x'}N^{s}(x')\left(\partial_s\phi\, \pi_{\phi}\right) \right\} \nonumber \\
&\equiv &\mathcal{P}_{1}+\mathcal{P}_{2}\,,
\label{spezzodue}
\end{eqnarray}
\end{widetext}
where the previous expression splits into the sum of the two Poisson bracket in virtue of the definition \eqref{PoissonBra}.

The first Poisson bracket $\mathcal{P}_{1}$ can be rewritten, after integration by part of the co-variant derivative in the second integral, in the following form

\begin{widetext}
\begin{equation}
\mathcal{P}_{1}=\left\{\int d^{3}x f(x)\left({h}_{ij}{\pi}^{ij}\right),\int d^{3}{x'}\left(2 D_a N^{s} h_{sb}(x') \pi^{ab}(x')\right)\right\}\equiv \mathcal{P}_{1a}+\mathcal{P}_{1b}+\mathcal{P}_{1c} \,.
\label{prima}
\end{equation}
\end{widetext}
It is easy to see that this Poisson bracket generates three terms: $\mathcal{P}_{1a}$ originated by the (non-null) Poisson Bracket of $h_{ij}(x)$ with $\pi^{ab}(x')$, $\mathcal{P}_{1b}$ by $\pi^{ij}(x)$ with $h_{sb}(x')$ and, finally, $\mathcal{P}_{1c}$ by $\pi^{ij}(x)$ with the metric terms in the co-variant derivative $D_aN^s(x')$. 

The first two terms are: 
\begin{widetext}
\begin{equation}
\mathcal{P}_{1a}=\int d^{3}y \int d^{3}x f(x)\pi^{pq}\delta^{3}(x-y)\int d^{3}{x'}\left(2 D_p N^{s}(x')h_{sq}(x')\right)\delta^{3}(x'-y),
\label{primotermine}
\end{equation}
and
\begin{equation}
\mathcal{P}_{1b}=-\int d^{3}y \int d^{3}x f(x)h_{pq}\delta^{3}(x-y)\int d^{3}{x'}\left(2 D_a N^{p}(x')\pi^{aq}(x')\right)\delta^{3}(x'-y)\,.
\label{spezzoduec}
\end{equation}
and clearly $\mathcal{P}_{1a}+\mathcal{P}_{1b}=0$.
\end{widetext}

The three-dimensional co-variant derivative $D_aN^s(x')$ is 
\begin{equation}
D_{a}N^{s}=\partial_{a}N^{s}+\Gamma^{s}_{ak}N^{k}\,,
\label{covariant}
\end{equation}
variations respect to the three-dimensional metric tensor $h_{ij}$ affect the 3-dimensional Levi Civita connection $\Gamma^{s}_{ak}$ in the following way \cite{Menotti2017}
\begin{equation}
\delta \Gamma^{s}_{ak}=\frac{1}{2}h^{sl}\left(D_k\delta h_{al}+D_a\delta h_{lk}-D_l\delta h_{ak}\right)\,.
\label{connectionvariation}
\end{equation}
Multiplying the previous expression with the quantity $N^k \pi^{at}h_{st}$, we get
\begin{equation}
\delta \Gamma^{s}_{ak} N^k \pi^{at}h_{st} = \frac{N^k}{2}\pi^{al}D_k \delta h_{al}\,.
\label{finalfor}
\end{equation}
\begin{widetext}
At this point we can easily see that $\mathcal{P}_{1c}$, the third term of \eqref{prima}, is nothing else but the Poisson bracket between $\pi^{ij}(x)$ and  the co-variant derivative $D_a N^s(x')$  

\begin{equation}
\mathcal{P}_{1}=\mathcal{P}_{1c}=\int d^{3}y \int d^{3}x f(x)h_{pq}\delta^{3}(x-y)\int d^{3}{x'} D_k\left(N^{k}{\pi}^{pq}\ \right)\delta^{3}(x'-y)\,.
\label{terzopezzo}
\end{equation}
The term $\mathcal{P}_{2}$ can be re-written as
\begin{eqnarray}
\mathcal{P}_{2}&=&-\int d^{3}y \int d^{3}x f(x)\pi_{\phi}\delta^{3}(x-y)\int d^{3}{x'} N^{a}(x')\partial_{a}\phi \delta^{3}(x'-y)\nonumber\\
&&-\int d^{3}y \int d^{3}x f(x)\phi\delta^{3}(x-y)\int d^{3}{x'}D_a\left( N^{a}(x')\pi_{\phi}(x')\right)\delta^{3}(x'-y)\,.
\label{partesuphi}
\end{eqnarray}
%\end{widetext}
Collecting $\mathcal{P}_{1}$ and $\mathcal{P}_{2}$ terms we obtain this expression for the Poisson bracket \eqref{spezzodue}
%\begin{widetext}
\begin{eqnarray}
\left\{\int d^3x f(x) C_{\phi}(x),\int d^3 x' N^i(x'){\mathcal{H}_{i}^{(-3/2)}}(x')\right\}&=& \int d^{3}y f(y) D_k \left( N^{k}(\pi_{h}-\phi\pi_{\phi}) \right) \nonumber\\
&=& -\int d^{3}y D_k f(y) N^{k}\left(\pi_{h}-\phi\pi_{\phi}\right)\, \approx 0\,.
\label{derivoprimario2app}
\end{eqnarray}
\end{widetext}
where we have simply integrated by parts.  

A further step toward the preservation, along the dynamics generated by the total Hamiltonian $H_T$, of the constraint $C_\phi$ is the calculation of the second term of \eqref{vincconf}, 
writing $C_{\phi}(x)$ and ${\mathcal{H}}^{(-3/2)}(x')$ in explicit form:
\begin{widetext}
\begin{eqnarray}
&&\left\{\int d^{3}x f(x)C_{\phi}(x), \int d^{3}x' N(x'){\mathcal{H}^{(-3/2)}}(x')\right\}\,\nonumber\\
&=& \Bigg\{\int d^{3}x f(x)\left({h}_{ij}{\pi}^{ij}-\phi\pi_{\phi}\right),
 \nonumber \\
&& \int d^{3}x'\sqrt{h}N(x')\bigg[ -\phi\;  {}^{3}R+\frac{1}{\phi h}\left( \pi^{ij}\pi_{ij}-\frac{{\pi_h}^2}{2}\right)
- \frac{3}{2\phi}D_i\phi D^i\phi +2D^iD_i\phi +U(\phi)  \bigg]\Bigg\}\nonumber\\
%\, .
&=& \mathcal{P}_{A1}+\dots+\mathcal{P}_{A6}+\mathcal{P}_{B1}+\dots+\mathcal{P}_{B5}\,.
\label{parentelunga}
\end{eqnarray}
\end{widetext}
The bi-linearity and Leibniz's rule for Poisson brackets  \cite{Menotti2017} grant 
%\eqref{parentelunga} can be decomposed 
breaking up
in two groups (A and B) of terms originated, roughly speaking, by the Poisson brackets of ${h}_{ij}{\pi}^{ij}$ and $-\phi\pi_{\phi}$ with the integral of the Hamiltonian constraint.
Therefore, labelling all terms, we have, for the first group A:\\
\begin{widetext}
\begin{eqnarray}
\mathcal{P}_{A1}&\equiv&\Bigg\{\int d^{3}x f(x)\left({h}_{ij}{\pi}^{ij}\right),
\int d^{3}x'\sqrt{h}N(x')\Bigg\}\bigg[ -\phi\;  {}^{3}R+\frac{1}{\phi h}\left( \pi^{ij}\pi_{ij}-\frac{{\pi_h}^2}{2}\right) \nonumber \\
&&-\frac{3}{2\phi}D_i\phi D^i\phi 
+2D^iD_i\phi                                     
 +U(\phi)  \bigg]\, ; \label{interA}\\
\mathcal{P}_{A2}&\equiv&\left\{\int d^{3}x f(x)\left({h}_{ij}{\pi}^{ij}\right),\int d^{3}{x'}N(x')\left(-\phi\;{}^{3}R_{ij}h^{ij}\right) \right\}\sqrt{h}\,
\label{intermezzo0}\,;\\
 \mathcal{P}_{A3}&\equiv&\Bigg\{\int d^{3}x f(x)\left({h}_{ij}{\pi}^{ij}\right),
\int d^{3}xN(x')\frac{1}{\phi h} \Bigg\}\sqrt{h}\left( \pi^{ij}\pi_{ij}-\frac{{\pi_h}^2}{2}\right)\,;
 \label{parentelonga01}\\
 \mathcal{P}_{A4}&\equiv& \Bigg\{\int d^{3}x f(x)\left({h}_{ij}{\pi}^{ij}\right),
\int d^{3}xN(x')\left( \pi^{ij}\pi_{ij}-\frac{{\pi_h}^2}{2}\right)\Bigg\}\sqrt{h}\frac{1}{\phi h} \,;
 \label{parentelonga02}\\
 \mathcal{P}_{A5}&\equiv& \Bigg\{\int d^{3}x f(x)\left({h}_{ij}{\pi}^{ij}\right),
\int d^{3}xN(x')\left(-\frac{3}{2\phi}D_i\phi D^i\phi  \right)\Bigg\}\sqrt{h}\,;
 \label{parentelonga03}\\
 \mathcal{P}_{A6}&\equiv&\Bigg\{\int d^{3}x f(x)\left({h}_{ij}{\pi}^{ij}\right),
\int d^{3}xN(x')\left(2D^iD_i\phi \right)\Bigg\}\sqrt{h}\,.
 \label{parentelonga04}
\end{eqnarray}
%\end{widetext}
The terms of the second group B are:
%\begin{widetext}
\begin{eqnarray}
\mathcal{P}_{B1}&\equiv& \left\{\int d^{3}x f(x)\left(\phi\pi_{\phi}\right),\int d^{3}{x'}\sqrt{h}N(x')\left(\phi\;\,{}^{3}R_{ij}g^{ij}\right) \right\}\,
\label{intermezzo0B}\,;\\
\mathcal{P}_{B2}&\equiv&-\Bigg\{\int d^{3}x f(x)\left(\phi\pi_{\phi}\right),
\int d^{3}xN(x')\frac{1}{\phi h} \Bigg\}\sqrt{h}\left( \pi^{ij}\pi_{ij}-\frac{{\pi_h}^2}{2}\right)\,;
 \label{parentelonga01B}\\
\mathcal{P}_{B3}&\equiv&\Bigg\{\int d^{3}x f(x)\left(\phi\pi_{\phi}\right),
\int d^{3}xN(x')\left(\frac{3}{2\phi}D_i\phi D^i\phi  \right)\Bigg\}\sqrt{h}\,;
 \label{parentelonga03B}\\
\mathcal{P}_{B4}&\equiv&-\Bigg\{\int d^{3}x f(x)\left(\phi\pi_{\phi}\right),
\int d^{3}xN(x')\left(2D^iD_i\phi \right)\Bigg\}\sqrt{h}\,;
 \label{parentelonga04B}\\
\mathcal{P}_{B5}&\equiv&-\Bigg\{\int d^{3}x f(x)\left(\phi\pi_{\phi}\right),
\int d^{3}xN(x')\left(U(\phi)\right)\Bigg\}\sqrt{h}\,.
 \label{parentelonga05B}
\end{eqnarray}
%\end{widetext}
We notice $\delta\sqrt{h}=\frac{1}{2}h^{ij}\delta h_{ij}$ in in \eqref{interA} (cfr. \cite{Menotti2017}), then
%\begin{widetext}
\begin{equation}
\Bigg\{\int d^{3}x f(x)\left({h}_{ij}{\pi}^{ij}\right),
\int d^{3}x'\sqrt{h}N(x')\Bigg\}
=\int d^{3}yf(y)N(y)(-\frac{3}{2})\,. 
\label{primaparente}
\end{equation}
\end{widetext}
and so the term $\mathcal{P}_{A1}$ is:
\begin{equation}
\mathcal{P}_{A1}=
%\bigg[ -\phi{}^{3}R+\frac{1}{\phi h}\left( \pi^{ij}\pi_{ij}-\frac{{\pi_h}^2}{2}\right) 
%\nonumber \\
%-\frac{3}{2\phi}D_i\phi D^i\phi 
%+2D^iD_i\phi                                     
% +U(\phi)  \bigg]
 \int d^{3}yf(y)N(y)(-\frac{3}{2}{\mathcal{H}}^{(-3/2)}(y))\,. 
\end{equation}

The term $\mathcal{P}_{A2}$ \eqref{intermezzo0} can be calculated in two steps. 
First we notice that  the variation $\delta h^{ij}=-h^{ik}h^{jh}\delta h_{kh}$. This entails that, 
%in the first step, one calculates the Poisson brackets of $\pi^{ij}h_{ij}$ with $h^{ij}$ in \eqref{intermezzo0}, providing the quantity
\begin{widetext}
\begin{equation}
\left\{\int d^{3}x f(x)\left({h}_{ij}{\pi}^{ij}\right),\int d^{3}{x'}N(x')\left(-\phi\;{}^{3}h^{ij}\right) \right\}R_{ij}\sqrt{h}=-\int d^{3}y f(y)N(y)\sqrt{h}\phi\, {}^{3}R\,.
\label{2pezzo}
\end{equation}
\end{widetext}
In the second step, we face the more delicate and difficult point of computing
\begin{equation}
    \left\{\int d^{3}x f(x)\left({h}_{ij}{\pi}^{ij}\right),\int d^{3}{x'}N(x')\left(-\phi\;{}^{3}R_{ij}\right) \right\}h^{ij}\sqrt{h}\,.
\end{equation}
Following \cite{Menotti2017}
\begin{widetext}
\begin{equation}
\int d^{3}x \left(-\sqrt{h}N\phi(\delta\,\, {}^{3}R_{ij})g^{ij}\right)=-\int d^{3}x \sqrt{h}\delta h_{ij}\left(D^{i}D^{j}-h^{ij}D^{m}D_{m}\right)\left(N\phi\right)\,,
\label{Rvar}
\end{equation}
%\end{widetext}
therefore 
%\begin{widetext}
\begin{equation}
\left\{\int d^{3}x f(x)\left({h}_{ij}{\pi}^{ij}\right),\int d^{3}{x'}N(x')\left(-\phi\;{}^{3}R_{ij}\right) \right\}h^{ij}\sqrt{h}=\int d^{3}y f(y) \sqrt{h} \left(-2 D_iD^i (N\phi)\right)\,.
\label{risRfin}
\end{equation}
%\end{widetext}
Finally we get for Eq. \eqref{intermezzo0}
%\begin{widetext}
\begin{equation}
\mathcal{P}_{A2}=-\int d^{3}y f(y)N(y)\sqrt{h}\phi\, {}^{3}R
-\int d^{3}y f(y) \sqrt{h} \left(2 D_iD^i (N\phi)\right)
\label{intermezzoIIA}\,.
\end{equation}
\end{widetext}

It is not difficult to check that the calculation of the term $\mathcal{P}_{A3}$, 
%
%\begin{equation}
%\Bigg\{\int d^{3}x f(x)\left({h}_{ij}{\pi}^{ij}-\phi\pi_{\phi}\right),
%\int d^{3}x'\frac{1}{\phi h}N(x')\Bigg\}\bigg[\sqrt{h}\left( \pi^{ij}\pi_{ij}-\frac{{\pi_h}^2}{2}\right)\bigg]
% \label{parentelonga}
%\end{equation}
%
in which we consider only the Poisson brackets between $\pi^{ij}$ and the inverse of the determinant $h$,  gives the result
\begin{equation}
\mathcal{P}_{A3}=\int d^{3}yf (y)\sqrt{h}N(y)\bigg[\frac{3}{\phi h}\left( \pi^{ij}\pi_{ij}-\frac{{\pi_h}^2}{2}\right)\bigg]\,.
\label{parentepiccola}
\end{equation}

One can easily see that the Poisson brackets $\mathcal{P}_{A4}=0$ as result
\begin{widetext}
\begin{equation}
\mathcal{P}_{A4}=\Bigg\{\int d^{3}x f(x)\left({h}_{ij}{\pi}^{ij}\right),
\int d^{3}x' N(x')\left( \pi^{ij}\pi_{ij}-\frac{{\pi_h}^2}{2}\right)\Bigg\}\sqrt{h}\frac{1}{\phi h}=0 \,.
\label{parentelonga02A}
\end{equation} 
\end{widetext}

The term $\mathcal{P}_{A5}$ can be re-written in the following way 
\begin{widetext}
\begin{equation}
\mathcal{P}_{A5}=\Bigg\{\int d^{3}x f(x){h}_{ij}{\pi}^{ij},
\int d^{3}x'\sqrt{h}N(x')\bigg(-\frac{3}{2}D_a\phi D_b\phi h^{ab}\bigg)\Bigg\}\,,
\label{como}
\end{equation}
\end{widetext}
and, in virtue of the observation made in the calculation for $\mathcal{P}_{A2}$ term regarding the variation of $\delta h^{ij}$, it gives the result
\begin{equation}
 \mathcal{P}_{A5}=\int d^{3}y f(y) \sqrt{h}N(y)\bigg(-\frac{3}{2}D_c\phi D^c \phi\bigg)\,.
 \label{ancora}
\end{equation}

Now we pass to calculate $\mathcal{P}_{A6}$ term:
%equivalent to the following Poisson bracket VI-A 
%\begin{equation}
%\Bigg\{\int d^{3}x f(x){h}_{ij}{\pi}^{ij},
%\int d^{3}x'\sqrt{h}N(x')\bigg(2D_c D^c\phi\bigg)\Bigg\}\,, 
%\label{pinno}
%\end{equation}
$D_i D^i\phi$ has a $h^{ij}$ coefficient in the contraction of the co-variant derivatives 
$D_i D^i\phi=D_a D_b h^{ab} \phi$, and in the second co-variant derivative on $\phi$:
$D_a D_b \phi=D_a (\partial_b \phi)=\partial_a \partial_b \phi-\Gamma^{s}_{ab}\partial_s \phi$. 
First we note that the variation of $\delta_h h^{ab}$ generates 
\begin{widetext}
\begin{equation}
\Bigg\{\int d^{3}x f(x){h}_{ij}{\pi}^{ij},
\int d^{3}x'\sqrt{h}N(x') h^{ab}\Bigg\} \bigg(2D_a D_b\phi\bigg)
=\int d^{3}y f(y) \sqrt{h}N(y)\bigg(2D_c D^c \phi\bigg)\, .
\label{primopezzo}
\end{equation}
\end{widetext}
We evaluate now the second term:
\begin{equation}
\Bigg\{\int d^{3}x f(x){h}_{ij}{\pi}^{ij},
\int d^{3}x'\sqrt{h}N(x') \bigg(2D_a D_b\phi\bigg)\Bigg\} h^{ab}\,,    
\end{equation}
remembering that variation $\delta_h$ of the term $D_a D_b \phi$ is:
\begin{equation}
\delta_h (D_a D_b \phi)=-\delta_h \Gamma^{s}_{ab}\partial_s \phi\, ,
\label{svario}
\end{equation}
and using the formula \eqref{connectionvariation} one gets 
\begin{equation}
h^{ab}\delta_h (D_a D_b \phi)=-\frac{1}{2}h^{ip}\left(2D_a \delta h_{pb}-D_p\delta  h_{ab}\right)h^{ab}D_i \phi \,\,.
\label{varioduecov}
\end{equation}
After integration by parts, the Poisson brackets generated by the presence of $h_{ij}$  in the connection $\Gamma$, are: 
\begin{widetext}
\begin{eqnarray}
&&\Bigg\{\int d^{3}x f(x){h}_{ij}{\pi}^{ij},
\int d^{3}x'\sqrt{h}N(x') \bigg(2D_a D_b\phi\bigg)\Bigg\} h^{ab}\nonumber\\
&&=-\int d^{3}y\sqrt{h} f(y)\left(2h^{kp}h^{lb}\delta ^{i}_{p} \delta ^{j} _{b} - h^{kl}h^{ab} \delta ^{i}_{a} \delta ^{j} _{b} \right)h_{ij}D_{l}(ND_k \phi)\nonumber\\
&&=\int d^{3}y\sqrt{h} f(y)D^{l}(ND_l \phi)\, . 
\label{risulti}
\end{eqnarray}
%\end{widetext}
Summing the two terms:
%\begin{widetext}
\begin{equation}
\mathcal{P}_{A6}=\int d^{3}y f(y) \sqrt{h}N(y)\left(2D_c D^c \phi\right)+\int d^{3}y\sqrt{h} f(y)D^{l}\left(N(y)D_l \phi\right)\, .
\label{risulto}
\end{equation}
\end{widetext}

Now we pass to the calculation of the second group (B) of Poisson brackets. It is straightforward that the following Poisson brackets $\mathcal{P}_{B1}$ \eqref{intermezzo0B} give this result
\begin{widetext}
\begin{equation}
\mathcal{P}_{B1}=-\int d^{3}y \sqrt{h}f(y)N(y)\phi\,\,{}^{3}R\,,
\end{equation}
and the term $\mathcal{P}_{B2}$ \eqref{parentelonga01B} gives
\begin{equation}
\mathcal{P}_{B2}=-\int d^{3}y\sqrt{h}\frac{N(y)f(y)}{\phi h}\left( \pi^{ij}\pi_{ij}-\frac{{\pi_h}^2}{2}\right)\,.
\label{pripo}
\end{equation}
Now we want to calculate the term $\mathcal{P}_{B3}$  \eqref{parentelonga03B}; 
we separate it in two terms, first we evaluate:

\begin{equation}
\Bigg\{\int d^{3}x f(x)\phi {\pi}_{\phi},
\int d^{3}x'\frac{3}{2}\sqrt{h}\frac{N(x')}{\phi}\Bigg\} D_c \phi D^c \phi
=\int d^{3}yf(y)\frac{3}{2}\sqrt{h}\frac{N(y)}{\phi}D_c \phi D^c \phi \,.
\label{primo1}
\end{equation}
\end{widetext}
\begin{widetext}
The contribution for the remaining term, after integration by parts, becomes

\begin{eqnarray}
-\Bigg\{\int d^{3}x f(x)\phi {\pi}_{\phi},
\int d^{3}x'D^c \left(\frac{3}{2}\sqrt{h}\frac{N(x')}{\phi}D_c \phi \right) 2\phi\Bigg\}
= \int d^{3}y f(y) \sqrt{h}\left(3D^c \left(N(y)D_c \phi\right)- 3\frac{N(y)}{\phi}D_c \phi D^c \phi\right).\nonumber\\
\end{eqnarray}
\end{widetext}
Adding the two terms  we arrive to the result that the Poisson bracket $\mathcal{P}_{B3}$ is 
\begin{widetext}
\begin{eqnarray}
\mathcal{P}_{B3}=\int d^{3}y f(y) \sqrt{h}\left(3D^c \left(N(y)D_c \phi\right)- \frac{3}{2}\frac{N(y)}{\phi} D_c \phi D^c \phi\right).
\label{final}
\end{eqnarray}
\end{widetext}
The last involved passage of these series of calculations is the evaluation of the term $\mathcal{P}_{B4}$ \eqref{parentelonga04B} the Poisson Brackets.
Double integration by parts gives 
\begin{widetext}
\begin{equation}
\mathcal{P}_{B4}=-\Bigg\{\int d^{3}x f(x)\phi {\pi}_{\phi},
2\int d^{3}x'\sqrt{h}D_c D^c N \phi\Bigg\}=2\int \sqrt{h} \phi (D^c D_c N) f(y) d^{3}y \,.
\label{doppiointegro}
\end{equation}
%\end{widetext}
Finally the term $\mathcal{P}_{B5}$ \eqref{parentelonga05B}, which is an immediate calculation 
%\begin{widetext}
\begin{equation}
\mathcal{P}_{B5}=-\Bigg\{\int d^{3}x f(x)\phi {\pi}_{\phi},
\int \sqrt{h} N U(\phi) d^{3}y \Bigg\}=\int d^{3}y f(y)N \phi \frac{d U(\phi)}{d \phi}. 
\label{finale2}
\end{equation}
%\end{widetext}
We are now ready to collect all the pieces for the calculation of the Poisson Brackets \eqref{parentelunga}
%\begin{widetext}
\begin{eqnarray}
&&\left\{\int d^{3}x f(x)C_{\phi}(x), \int d^{3}x' N(x'){\mathcal{H}}^{(-3/2)}(x')\right\} \nonumber \\
&=&\int d^{3}y \Bigg[f(y) N \left(-\frac{3}{2}{\mathcal{H}^{(-3/2)}(y)} \right)
-\sqrt{h}f(y)N\phi\,\,{}^{3}R -2 f(y) \sqrt{h} \left(D_iD^i (N\phi)\right)\nonumber \\
 && +\frac{3\sqrt{h} N f(y)}{\phi h}\left( \pi^{ij}\pi_{ij}-\frac{{\pi_h}^2}{2}\right)
 -\frac{3}{2}\sqrt{h}\frac{N}{\phi}D_c \phi D^c \phi f(y)
  +2 f(y)\sqrt{h} N (D^c D_c \phi)\nonumber\\
 && + \sqrt{h}f(y)D^{c}(ND_c \phi)-\sqrt{h}f(y) N \phi\,\,{}^{3}R-\frac{\sqrt{h} N f(y)}{\phi h}\left( \pi^{ij}\pi_{ij}-\frac{{\pi_h}^2}{2}\right)-\frac{3}{2}\sqrt{h}\frac{N}{\phi}D_c \phi D^c \phi f(y)\nonumber \\ 
 &&+3\sqrt{h}f(y)D^{c}(ND_c \phi) + 2f(y)\sqrt{h} \phi (D^c D_c N)+
  f(y)N \phi \frac{d U(\phi)}{d \phi}\,
  \Bigg].
 \label{infondo}
 \end{eqnarray}
\end{widetext}
Recalling that, from equation \eqref{eqstophi} in the case $\omega=-\frac{3}{2}$, 
\begin{widetext}
\begin{equation}
    \phi \frac{d U(\phi)}{d \phi}-2U(\phi)=0\,, 
    \label{consisto}
 \end{equation}
one finally finds 

\begin{equation}
 %\Bigg\{&&\int d^{3}x f(x)\left({h}_{ij}{\pi}^{ij}-\phi\pi_{\phi}\right),
%\int d^{3}x'\sqrt{h}N(x')\bigg[ -\phi\;  {}^{3}R+\frac{1}{\phi h}\left( \pi^{ij}\pi_{ij}-\frac{{\pi_h}^2}{2}\right) \nonumber \\
%&&- \frac{3}{2\phi}D_i\phi D^i\phi 
%+2D^iD_i\phi                                     
% +U(\phi)  \bigg]\Bigg\}
\left\{\int d^{3}x f(x)C_{\phi}(x), \int d^{3}x' N(x'){\mathcal{H}}^{(-3/2)}(x')\right\} 
=\frac{1}{2}\int d^{3}yN(y)f(y){\mathcal{H}}^{(-3/2)}(y)
\approx 0\,.
 \label{endin}
 \end{equation}
 \end{widetext}
 
Recapitulating - adding \eqref{derivoprimario2app}  and \eqref{endin} - we 
obtain the condition for the preservation of the primary constrain $C_{\phi}(x)$
\begin{widetext}
\begin{equation}
\left\{\int d^{3}x f(x)C_{\phi}(x),H_{T}^{(-3/2)}\right\}=-\int d^{3}y D_k f(y)  \left( N^{k}C_{\phi} \right)+\frac{1}{2}\int d^{3}yN(y)f(y){\mathcal{H}}^{(-3/2)}(y)\approx 0\,.
\label{vincsumm app}
\end{equation}
\end{widetext}

\subsection{Poisson bracket of the Hamiltonian constraint ${\mathcal{H}}^{(-3/2)}$ }
Since the momentum $\pi_{\phi}$ is absent in the Hamiltonian constraint \eqref{hamiltoconst}, our computations are restricted to the variation of the metric functions $h_{ij}$ and its conjugate momenta $\pi^{ij}$. Non-algebraic variations of the metric tensor are generated, as it can be easily see, by the trace of Ricci tensor ${}^{3} R$, as it can be easily see in the equation \eqref{Rvar}, and the the double covariant derivative of the scalar field $D^c D_c \phi$ as shown in the equations \eqref{svario} and \eqref{varioduecov}.
Therefore the previous Poisson brackets \eqref{possyhami} is equivalent to the calculation of the following integrals 
\begin{widetext}
\begin{eqnarray}
&&\left\{ \int d^{3}x N(x) {\mathcal{H}}^{(-3/2)}(x), \int d^{3}x N'(x') {\mathcal{H}}^{(-3/2)}(x')\right\}\nonumber\\
&=&\int d^{3}y\Bigg[ \int d^{3}x \sqrt{h}\bigg(-\frac{\delta h_{lm}(x)}{\delta h_{ij}(y)} \bigg) \left(D^{l}D^{m}-h^{lm}D^{k}D_{k}\right)(N\phi)  \\
&&\Bigg(\frac{\delta}{\delta \pi^{ij}(y)}\int d^{3}x'\frac{1}{\phi h}
\Big( \pi^{ab}\pi_{ab}
 -\frac{{\pi_h}^2}{2}\Big)\sqrt{h}N'(x')\Bigg)
\Bigg]- (N\leftrightarrow N') +\int d^{3}y \Bigg[\left(\frac{\delta h_{cd}(x)}{\delta h_{ij}(y)}\right)\int d^{3}x\sqrt{h}N(x)\nonumber \\
&&\bigg(2h^{kp}h^{lb}\delta ^{c}_{p} \delta ^{d} _{b} 
-h^{kl}h^{ab} \delta ^{c}_{a} \delta ^{d} _{b} \bigg)D_{l}(N(x) D_k \phi)\Bigg(\frac{\delta} {\delta \pi^{ij}(y)}\int d^{3}x'N'(x')\sqrt{h}\frac{1}{\phi h}\Big( \pi^{ab}\pi_{ab}
 -\frac{{\pi_h}^2}{2}\Big)\Bigg)\Bigg]- (N \leftrightarrow N')\, \nonumber
 \label{maestosa}
\end{eqnarray}
\end{widetext}
which, performing the calculations, reduces to
\begin{widetext}
\begin{eqnarray}
&&\left\{ \int d^{3}x N(x) {\mathcal{H}}^{(-3/2)}(x), \int d^{3}x N'(x') {\mathcal{H}}^{(-3/2)}(x')\right\} \nonumber \\
&=&\int d^{3}y\Bigg[-\Bigg((D^i D^j-h^{ij}D^k D_k)(N\phi)\Bigg)\frac{N'}{\phi}(2\pi_{ij}-h_{ij}\pi_h)\Bigg]-(N \leftrightarrow N')\nonumber \\
&&+2\int d^3 y (D^i N)(D^j \phi)\frac{N'}{\phi}(2\pi_{ij}-h_{ij}\pi_h) -(N\leftrightarrow N')+\int d^3 y (D^k N)(D_k \phi)\frac{N'}{\phi}(\pi_h)-(N\leftrightarrow N')\,.
\label{racco}
\end{eqnarray}
\end{widetext}
Further, we compute the double derivatives on $N\phi$ and discard the term $D^k D_k \phi$ since it cancels with its similar term in $N\leftrightarrow N'$, 
\begin{widetext}
\begin{eqnarray}
&&\left\{ \int d^{3}x N(x) {\mathcal{H}}^{(-3/2)}(x), \int d^{3}x N'(x') {\mathcal{H}}^{(-3/2)}(x')\right\} \nonumber \\
&=&-\int d^{3}y\Bigg[(D^i D^j)(N)\phi(\frac{N'}{\phi})(2\pi_{ij})-(D^k D_k)(N)(\frac{N'}{\phi})
(\pi_h)+2(D^i N)(D^i \phi)(\frac{N'}{\phi})(2\pi_{ij})\nonumber \\
&&-2(D^k N)(D_k \phi))(\frac{N'}{\phi})(\pi_h)+(D^k D_k)(N)\phi(\frac{N'}{\phi})(\pi_h)+
2(D^k N)(D_k \phi))(\frac{N'}{\phi})(\pi_h)\Bigg]\nonumber \\
&&-(N\leftrightarrow N') +2\int d^{3}y(D^i N)(D^j \phi)(\frac{N'}{\phi})(2\pi_{ij}-h_{ij}\pi_h))-(N\leftrightarrow N') \nonumber \\
&&+\int d^{3}y (D^k N)(D_k \phi))(\frac{N'}{\phi})(\pi_h) -(N\leftrightarrow N')\,.
\end{eqnarray}
\end{widetext}
Simplifying, we get
\begin{widetext}
\begin{eqnarray}
&&\left\{ \int d^{3}x N(x) {\mathcal{H}}^{(-3/2)}(x), \int d^{3}x N'(x') {\mathcal{H}}^{(-3/2)}(x')\right\} \nonumber \\
&&=\Bigg[-\int d^{3} y(D^i D^j)(N)(2 \pi_{ij})N' -\int d^{3}y(D^i N)(D_i \phi)\frac{N'}{\phi}(\pi_{h})\Bigg] -(N\leftrightarrow N') \,.
\label{simplio}
\end{eqnarray}
\end{widetext}
Integrating by parts, the equation above becomes
\begin{widetext}
\begin{eqnarray}
&&\left\{ \int d^{3}x N(x) {\mathcal{H}}^{(-3/2)}(x), \int d^{3}x N'(x') {\mathcal{H}}^{(-3/2)}(x')\right\} \nonumber \\
&&=\int d^{3}y(ND^{i}N' - N'D^{i}N)(-2D^{j}\pi_{ij} + \pi_{\phi}\partial_{i}\phi - \pi_{\phi}\partial_{i}\phi)%\nonumber \\
+\int d^{3}y (ND^{i}N' - N'D^{i}N)(D_i \log \phi)(\pi_{h})\,.
\end{eqnarray}
\end{widetext}

At the end, we find that the Dirac's constraint algebra generated by the Poisson brackets of the Hamiltonian constraint \eqref{possyhami} is 
\begin{widetext}
\begin{eqnarray}
&&\left\{ \int d^{3}x N(x) {\mathcal{H}^{(-3/2)}}(x), \int d^{3}x N'(x') {\mathcal{H}^{(-3/2)}}(x')\right\} \nonumber \\
&&=\int d^{3}y(ND^{i}N'- N'D^{i}N){\mathcal{H}_i}
+\int d^{3}y(ND^{i}N'- N'D^{i}N)(D_i \log\phi)C_{\phi}\approx 0\,. 
\label{finioAPP}
\end{eqnarray}
\end{widetext}

\bibliography{bransdickepartcase}

\end{document}